\documentclass[a4paper,11pt]{article}
\usepackage{pos}

\usepackage[dvipsnames]{xcolor}
\usepackage[utf8]{inputenc}
\usepackage[bbgreekl]{mathbbol}
\usepackage{multicol}
\usepackage{float} 
\usepackage{dsfont}

\usepackage{extarrows}
\usepackage{xcolor}
\usepackage{upgreek}
\usepackage{xfrac}
\usepackage[inline]{enumitem}

\setlist{nolistsep}
\usepackage[french, italian, english]{babel}
\usepackage[all]{xy}

\usepackage{newtxtext}
\usepackage{newtxmath}

\usepackage{booktabs}
\usepackage{caption}
\usepackage{dsfont}
\usepackage{mathtools}
\usepackage{slashed}

\usepackage[makeroom]{cancel}

\usepackage[title]{appendix}

\usepackage{tikz}
\usepackage{tikz-cd}
\usetikzlibrary{cd}
\tikzcdset{
arrow style=tikz,
diagrams={>={Straight Barb[scale=0.8]}}
}
\usetikzlibrary{matrix,arrows,decorations.pathmorphing}

\DeclareMathAlphabet{\mathpzc}{OT1}{pzc}{m}{it}

\usepackage{fancybox}

\usepackage{mathrsfs}

\usepackage{scrextend}

\makeatletter
\renewcommand*\env@matrix[1][\arraystretch]{%
  \edef\arraystretch{#1}%
  \hskip -\arraycolsep
  \let\@ifnextchar\new@ifnextchar
  \array{*\c@MaxMatrixCols c}}
\makeatother

\newcommand{\defeq}{\vcentcolon=}

\newcommand\M{\mathcal{M}}

\newcommand\RR{\mathbb{R}}
\newcommand\CC{\mathbb{C}}

\newcommand\C{\mathcal{C}}
\newcommand\id{\textit{id}}
\newcommand\T{\mathcal{T}}
\newcommand\G{\mathcal{G}}
\renewcommand\H{\mathcal{H}}

\newcommand\E{\mathcal{E}}

\renewcommand\L{\mathcal{L}}
\renewcommand\S{\mathcal{S}}

\newcommand\U{\mathcal{U}}
\newcommand\SO{\mathcal{SO}}
\newcommand\SL{\mathcal{SL}}

\newcommand\K{\mathcal{K}}
\newcommand\J{\mathcal{J}}
\renewcommand\O{\mathcal{O}}

\newcommand\D{\mathcal{D}}

\newcommand\vphi{\varphi}

\renewcommand\l{\text{\tiny{L}}}

\renewcommand\epsilon{\varepsilon}

\newcommand\rarrow{\rightarrow}

\newcommand\diff{\mathfrak{diff}}

\newcommand\LieG{\mathfrak{g}}
\newcommand\LieH{\mathfrak{h}}

\renewcommand\b{\bar }

\renewcommand\d{\partial}
\newcommand\s{\sigma}
\newcommand\bs{\boldsymbol}

\renewcommand\-{^{-1}}

\newcommand\ad{\text{ad}}
\renewcommand\id{\text{id}}

\makeatletter

\newcommand{\Rmnum}[1]{\expandafter\@slowromancap\romannumeral #1@}
\makeatother

\makeatletter
\newcommand{\leqnomode}{\tagsleft@true\let\veqno\@@leqno}
\newcommand{\reqnomode}{\tagsleft@false\let\veqno\@@eqno}
\makeatother

\DeclareMathOperator{\Diff}{Diff}

\DeclareMathOperator{\Tr}{Tr}

\theoremstyle{definition}



\title{Lecture Notes on Symmetry Reduction via the Dressing Field Method}

\author*[a,b,c]{Lucrezia Ravera}

\affiliation[a]{DISAT, Politecnico di Torino -- PoliTO, \\
Corso Duca degli Abruzzi 24, 10129 Torino, Italy.}

\affiliation[b]{Istituto Nazionale di Fisica Nucleare, Section of Torino -- INFN, \\
Via P. Giuria 1, 10125 Torino, Italy.}

\affiliation[c]{Grupo de Investigación en Física Teórica (GIFT), \\ Universidad Cat\'{o}lica De La Sant\'{i}sima Concepci\'{o}n -- UCSC, \\
Alonso de Ribera 2850, Concepción, Chile.}

\emailAdd{lucrezia.ravera@polito.it}

\abstract{These notes -- prepared for the conference school \emph{Foundations of General-Relativistic Gauge Field Theory}, held on March 17-19, 2026 at the Politecnico di Torino -- present introductory material on symmetry reduction in general-relativistic Gauge Field Theory (gRGFT) via the Dressing Field Method (DFM).
The DFM provides a systematic framework for extracting gauge- and diffeomorphism-invariant,  manifestly relational, physical observables and degrees of freedom in gRGFT.
A range of illustrative examples are discussed, spanning both Gauge Field Theory and general-relativistic settings. 
These include applications to non-Abelian Chern-Simons theory, Maxwell electromagnetism, the non-Abelian Higgs model, supersymmetric field theory, General Relativity, and scalar coordinatization.}

\FullConference{Proceedings of the Conference School ``Foundations of General-Relativistic Gauge Field Theory'' (FGRGFT2026) \\ 17 March - 19 March, 2026 \\ Politecnico di Torino, Torino, Italy
}

\tableofcontents

\begin{document}
\maketitle

\section{Introduction}
\label{Introduction}

The two pillars of modern theoretical physics -- the Standard Model (SM) and General Relativity (GR) -- are both grounded in the principle of local symmetries. The former is a Gauge Field Theory (GFT): it admits a classical Lagrangian formulation, but it is also a quantum field theory, characterized by internal local symmetries known as \emph{gauge symmetries}. The latter, by contrast, is a classical theory (as no complete and fully developed theory of quantum gravity currently exists), whose local symmetries are given by the \emph{diffeomorphisms} of the ``spacetime manifold'' $M$ -- notion itself that, as we shall see, will be refined during these lectures.\footnote{For a total of 2.5 hours of in-person lectures.}

The framework that emerges from combining these two theories, at least at the classical level, is that of \emph{general-relativistic Gauge Field Theory} (gRGFT). 
Their local symmetries -- which yield both symmetry principles and the principle of general covariance -- consist of gauge transformations together with diffeomorphisms. The full local symmetry group can be written as $\Diff(M) \ltimes \H$,
where $\H$ denotes the gauge group.

Physical observables (``Dirac observables'') and physical degrees of freedom (d.o.f.) must be invariant under both gauge transformations and diffeomorphisms. 
These are precisely the quantities accessible in experiments and observations. 
Consequently, a variety of techniques have been developed to achieve such invariance.

Among these, the BRST-BV formalism (named after Becchi-Rouet-Stora-Tyutin and Batalin-Vilkovisky) \cite{BRS-75,Becchi:1975nq,Tyutin:1975qk,Becchi:1996yh,Barnich-et-al2000,Batalin:1981jr}, which is based on the dynamical implementation of gauge fixing, and consequent quantization of a theory, 
and remains the most widely used and refined framework in high-energy physics. 
Other approaches involve symmetry reduction methods, including the FMS (Frölich-Morchio-Strocchi) \cite{Frohlich-Morchio-Strocchi80,Frohlich-Morchio-Strocchi81,Maas2020,Maas2023} approach and -- the main topic of these lectures -- the more recent \emph{Dressing Field Method} (DFM), see, e.g., \cite{Fournel:2012cr,Francois2023-a,JTF-Ravera2024gRGFT,JTF-Ravera2024-SUSY,JTF-Ravera2024ususyDFM,JTF-Ravera2025DFMSusyReview&Misu,JTF-Ravera2025offshellsusy,JTF-RaveraNoBdyPb2025,Francois:2025jro,Francois:2025ptj}.\footnote{Let us mention here that both the DFM and FMS approaches bypass the notion of spontaneous symmetry breaking, see, e.g., \cite{Francois2018,Berghofer-et-al2023}.}

The DFM will be the focus here because it is likely less familiar to the audience than other approaches, and moreover it provides a particularly clean and transparent \emph{relational} interpretation. 
Indeed, it implements at a technical level the so-called \emph{point-coincidence argument} formulated by Albert Einstein (as later termed by philosophers of physics), which is the reply to his own \emph{hole argument} (cf., e.g., \cite{Norton1987,Norton1988,Earman-Norton1987,EarmanNorton,Norton1993,Norton2000,Earman1989,Stachel2014,JTF-Ravera2024c,Berghofer:2025ius}), 
and constitutes a powerful tool for extracting the gauge- and diffeomorphism-invariant content -- that is, the true physical d.o.f. -- of a gRGFT.

In these lectures, I will focus exclusively on the \emph{classical} aspects (formulation and application of the DFM). 
The interested reader can find further developments and applications at the \emph{quantum} level in \cite{JTF-Ravera2024NRrelQM,Francois:2025lqn} (see also a side note in \cite{Lingua:2024tsn}).
Moreover, I will adopt a purely \emph{field-theoretic} and physical perspective, aiming for a presentation that is as pedagogical and clearly oriented toward the physics community as possible. 
I will also provide diverse applications of the DFM to GFT and general-relativistic physics.
Readers interested in a deeper understanding can explore the geometric aspects and the full potential of the DFM in the context of field space viewed as a bundle in differential geometric terms in \cite{JTF-Ravera2024gRGFT}.

The remainder of these notes is organized as follows: In Section \ref{Hole and point-coincidence arguments in a nutshell}, I provide a brief conceptual overview of the hole argument -- Einstein's own conceptual critique of General Relativity (GR) -- and of the point-coincidence argument developed in response to it, which enabled the fully fledged development of the theory of GR by Einstein himself.
This dialectic will allow us to appreciate the importance of \emph{relationality} as a key insight of general-relativistic physics. 
By ``relationality'', for short, we mean that physical entities (e.g., fields) dynamically co-define each other, without any non-dynamical fixed background structures; there are no non-coupled physically meaningful entities.
This conceptual notion is implemented technically through the DFM, which realizes, in a precise way, the point-coincidence argument extended to gRGFT, and naturally features a relational interpretation.

In Section \ref{Dressing Field Method for gauge symmetries}, we will address the formulation and application of the DFM to GFTs, and hence the systematic reduction of gauge symmetries in order to construct/achieve \emph{dressed}, that is, gauge-invariant, physical, and relational d.o.f./descriptions.
This method, let us recall, is based on a ``conditional statement'': \emph{If} a dressing field can be extracted (from the field content of the theory), then dressed objects can be constructed. 
These are automatically gauge-invariant (and diffeomorphism-invariant, in the case of general-relativistic physics) and manifestly relational.
I will first define the formalism at both the kinematical and dynamical levels, also explaining the differences between the DFM and gauge fixing. 
I will then provide several examples of applications, including the case of non-Abelian Chern-Simons (CS) theory; the extraction of a dressing field from a functional constraint on the fields in Maxwell electromagnetism (taking the Lorenz gauge condition as an example); the $\U(1)$ theory with a complex scalar field, from which we will construct the dynamics of the Abelian Higgs model and show that the DFM avoids the usual interpretation in terms of spontaneous symmetry breaking (SSB); and a simple example of ``dressing the Lorentz symmetry'' in, e.g., GR (recalling that Lorentz symmetry is a gauge symmetry in this general-relativistic setting).

I will then focus, in Section \ref{Dressing Field Method in supersymmetric field theory}, on the case of supersymmetric (gauge) field theory, showing how to extract a dressing field from the Rarita-Schwinger (RS) field (a procedure commonly referred to as gauge fixing in standard supersymmetry/supergravity). 
In this way, we obtain the RS field as a relational, self-dressed object, a singlet under supersymmetry. 
In the same section, I will introduce the BRST formalism at the algebraic level, along with its dressed version obtained via the DFM, and then apply it to the RS case (for extensions to supergravity, in particular in a geometric superspace approach, I refer the reader to \cite{JTF-Ravera2024-SUSY,JTF-Ravera2025DFMSusyReview&Misu}).
I will also present a recently introduced framework, known as \emph{Matter-Interaction Supergeometric Unification} (MISU), which is based on the DFM and is fundamentally relational. 
This framework allows one to include matter fields within a superconnection together with bosonic interaction fields, thus employing a supergeometric setting that does not require matching between bosonic and fermionic d.o.f. \cite{Sohnius:1985qm}. 
Here as well, I will discuss the dressed BRST formulation.

Finally, in Section \ref{Dressing Field Method for general-relativistic theories}, I will examine the DFM in the general-relativistic context, presenting both the kinematics and dynamics of dressing for diffeomorphisms. 
I will introduce the fundamental concept of \emph{dressed regions} (which allows one to resolve -- or, better, dissolve -- the so-called ``boundary problem''), and provide examples in GR with scalar fields, where the DFM enables a form of physical (scalar) coordinatization in general-relativistic physics. 
I will conclude in Section \ref{Concluding remarks} with some remarks on applications to Metric-Affine Gravity (MAG) and conformal gravity, where, in a general-relativistic setting, the focus is again on the reduction of gauge symmetries in these theories, as well as with comments on the quantum theory.

Appendix \ref{Stueckelberg trick vs the DFM} discusses the difference between the ``Stueckelberg trick'' and the DFM, while Appendix \ref{Dirac dressing} presents the Dirac dressing, for readers interested in further examples developed in the more or less recent literature.

Most of the material presented in these lecture notes is drawn from, and revisited from, work carried out in collaboration with J. François, as part of a broader research program based on the DFM that is still currently actively under development.

\section{Hole and point-coincidence arguments in a nutshell}\label{Hole and point-coincidence arguments in a nutshell}

In this section, we briefly revisit the logic of the hole argument and the point-coincidence argument in the context of GR. For their development within GFT, as well as their extension and generalization to the gRGFT framework, we refer the reader to, e.g., \cite{JTF-Ravera2024c,JTF-RaveraNoBdyPb2025,Berghofer:2025ius}.

We denote the generic field content of a general-relativistic theory by $\upphi$. 
The logic of the \emph{hole argument} is as follows. 
By assumption, the field equations $E(\upphi)=0$ of a GR theory are $\Diff(M)$-covariant. Hence, if $\upphi$ is a solution, then so is $\psi^*\upphi$ (which denotes the \emph{pullback} action on $\upphi$ of the diffeomorphism $\psi$) for any $\psi \in \Diff(M)$, since
$E(\psi^* \upphi) = \psi^* E(\upphi) = 0$.
Consider $\upphi$, $\upphi'$ belonging to the same $\Diff(M)$-orbit $\O_\upphi$, i.e., $\upphi'=\psi^*\upphi$, with $\psi$ a compactly supported diffeomorphism with support $D_\psi \subset M$ (the ``hole'').
We have $\upphi=\upphi'$ on $M/D_\psi$, but $\upphi \neq \upphi'$ on $D_\psi$.
As a result,
the field equations appear to admit an ill-defined Cauchy problem: they do not uniquely determine the evolution of $\upphi$ within $D_\psi$. \emph{Prima facie}, the theory therefore seems to exhibit a form of indeterminism.
A question therefore arises: How are deterministically evolving \emph{physical spatiotemporal d.o.f.} represented in GR?

The resolution was provided by Einstein himself through what was later termed the \emph{point-coincidence argument}, which asserts that the physical content of the theory (its observables) is fully exhausted by the ``pointwise coincidental values of fields'' $\upphi$, and is $\Diff(M)$-invariant.
That is, all solutions in $\O_\upphi$ represent the same physical state.
The deterministically evolving d.o.f., which we may denote $[\upphi]$, are \emph{not} within the individual (``purely mathematical'') fields $\upphi$, but instantiated by the $\Diff(M)$-invariant \emph{relations} among them.
As a consequence, also the points of $M$ are \emph{not} physical spatiotemporal events.
\emph{Physical spacetime} is represented in GR as the $\Diff(M)$-invariant network of relations among physical fields.

As we shall discuss in the following, since they allow one to achieve $\Diff(M)$-invariance relationally, with physical spatiotemporal field d.o.f. co-defining each other as coincidental values of fields,
the \emph{dressed variables} constructed with the DFM formally implement the \emph{point-coincidence argument}, and are immune to the \emph{hole argument}:\footnote{All this being  achieved already at the kinematical level via the DFM.}
Some of the spatiotemporal d.o.f. of the ``bare'' fields $\upphi$ are used to invariantly ``coordinatize'' 
the remaining spatiotemporal d.o.f. of $\upphi$; the dressed fields then 
represent the $\Diff(M)$-invariant relations co-defining the physical d.o.f. embedded in bare fields -- 
they are \emph{relational variables}.

\section{Dressing Field Method for gauge symmetries}\label{Dressing Field Method for gauge symmetries}

Via the DFM one produces gauge-invariant variables out of the field space $\Phi=\{\upphi\}$ of a general-relativistic Gauge Field Theory (gRGFT). We recall that the dressing procedure can be implemented, via the DFM, both in the case of internal gauge symmetries $\mathcal{H}$ and in the case of $\Diff(M)$, as we will discuss in these notes. 
Here we start by reviewing the application of the DFM to gauge symmetries, considering first the kinematics and then the dynamics.
Note that the DFM applies both at the \emph{finite} (\emph{non-perturbative}) and at the linearized/infinitesimal, perturbative levels.

\subsection{Dressed GFT kinematic}\label{Dressed GFT kinematic}

Consider a GFT with field content $\upphi=\lbrace{ A, \varphi \rbrace}$ -- e.g., Yang-Mills (YM) theory and QED -- on $M$, where $A$ is the 1-form gauge potential and $\varphi$ represents the matter field content (a collection of matter fields), both supporting the action of the gauge group 
\begin{align}
    \mathcal{H} := \lbrace{ \gamma: M \to H  \,| \, \gamma^\eta = \eta^{-1} \gamma \eta \rbrace} , \quad \gamma,\eta \in \H ,
\end{align}
so that they
gauge-transform as
\begin{align}
\label{gaugetrAvphi}
    A^\gamma:=  \gamma\- A \gamma + \gamma\- d\gamma, \quad \varphi^\gamma:= \gamma\- \varphi, \quad \gamma \in \H .
\end{align}
A \emph{dressing field} is a \emph{group-valued} field, a smooth map, 
\begin{align}
\label{dressing-field-101}
    u : M \rightarrow H , \quad \text{s.t.} \quad u^\gamma = \gamma\- u , \quad \gamma \in \H .
\end{align}
Note that, by definition, a dressing field must be both group-valued and transform under gauge transformations as in \eqref{dressing-field-101} in order to qualify as such and to be employed within the DFM.\footnote{This is also the reason why we cannot regard the DFM as a mere field redefinition, even though it involves a ``reshuffling'' of the bare d.o.f. to build composite, automatically gauge-invariant, dressed field variables. A simple ``field redefinition'' does not, in general, produce manifestly relational and invariant objects, whereas dressing does so.}
Moreover, a key aspect of the DFM is that a dressing field should always be extracted from the field content of the theory. 
This means that it has to be a \emph{field-dependent dressing field} $u=u[\upphi]$, so that $u^\gamma := u[\upphi^\gamma] = \gamma\- u [\upphi]$.

When such a dressing field is found/built, we can define the \emph{dressed fields} $\upphi^u = \lbrace{ A^u , \varphi^u \rbrace}$, which are given by
\begin{align}
\label{dressed-fields}
A^u:=  u\- A u + u\- du, \quad \varphi^u:= u\- \varphi .
\end{align} 
This illustrates the DFM ``rule of thumb'': To dress fields or functionals thereof, we compute first their gauge transformations, then formally substitute the gauge parameter $\gamma$ with the dressing field $u$. 
The resulting expressions are $\H$-invariant by construction -- which can also be easily checked explicitly.

Let us stress that, since $u \notin \H$, dressed fields are not gauge transformed fields, and, in particular, a dressing via the DFM is \emph{not} a gauge fixing, as explained in detail in, e.g., \cite{Berghofer-Francois2024,JTF-Ravera2025DFMSusyReview&Misu}.
When $u=u[\upphi]$ is a field-dependent dressing field, the DFM has a natural \emph{relational} interpretation \cite{JTF-Ravera2024c,JTF-Ravera2024gRGFT} (which a gauge fixing has not; moreover, a gauge fixing is not invariant): the dressed fields $\{A^{u[A, \vphi]}, \vphi^{u[A, \vphi]}\}$ in \eqref{dressed-fields} represent the \emph{gauge-invariant} \emph{relations} among the physical internal d.o.f. \mbox{embedded} in $\{A, \vphi\}$. 

Let us also mention that, analogously -- that is, using the DFM rule of thumb -- from the bare curvature and covariant derivative
\begin{align}
    F:= dA+\tfrac{1}{2}[A,A] , \quad D\varphi :=d\varphi + \rho_*(A)\varphi ,
\end{align}
gauge-transforming as 
\begin{align}
    F^\gamma = \gamma^{-1} F \gamma , \quad D \varphi^\gamma = \rho(\gamma)^{-1} D\varphi = d \varphi^\gamma + \rho_* (A^\gamma) \varphi^\gamma ,
\end{align}
and where $F$ obeys the Bianchi identity $D^A F=0$,
one can then derive the dressed curvature and dressed covariant derivative:
\begin{align}
    F^u= u^{-1}Fu = dA^u+\tfrac{1}{2}[A^u,A^u] , \quad D^u\varphi^u :=d\varphi^u + \rho_*(A^u)\varphi^u ,
\end{align}
which are automatically $\H$-invariant (and $F^u$ obeys the dressed Bianchi identity $D^{A^u}F^u=0$).

\subsection{Dressed dynamics in GFT}\label{Dressed dynamics in GFT}

The above pertain to the kinematics.
The dynamics of a GFT is specified by a Lagrangian $L=L(\upphi)$, that is typically required to be quasi-invariant under $\H$, i.e., 
$L(\upphi^\gamma)=L(\upphi) + db(\gamma;\upphi)$, for $\gamma \in \H$.
This guarantees the $\H$-covariance of the field equations $ E(\upphi)=0$ extracted from $L$. 

Given a dressing $u$, we  define the \emph{dressed \mbox{Lagrangian}} 
\begin{align}
\label{dressed-Lagrangian-int}
L(\upphi^u):=L(\upphi) + db(u;\upphi),
\end{align}
which is \emph{strictly} $\H$-invariant by construction (indeed, it is a functional of the dressed, $\H$-invariant fields $\upphi^u$; the right-hand side of \eqref{dressed-Lagrangian-int} shows the ``unfolding'' of the dressing in terms of bare variables) -- this is a case of the DFM rule of thumb. 
The field equations for the dressed fields, $ E(\upphi^u)=0$, have the \emph{same functional expression} as those of the ``bare'' fields for the bare theory: here, as can be seen when unfolding the dressing, they still differ by a boundary term (which may or may not involve non-localities, possibly induced by the presence of non-local dressing field, as we shall see in the examples), 
see \cite{JTF-Ravera2024gRGFT}.
The dressed field equations are \emph{deterministic}, they specify uniquely the evolution of the physical internal d.o.f. represented by the dressed fields \eqref{dressed-fields}.
One may check that $E = 0 \Rightarrow E^u=0$ \cite{JTF-Ravera2024gRGFT}. 
This
is the reason why the bare formalism gives sensible physical
results, as it is actually the relational version of
the theory which is confronted with experimental tests, not a
gauge-fixed version of it as is often said -- let us indeed insist that $L^u$ and $E^u$ are \emph{not} gauge-fixed versions of the bare $L$ and $E$.

\subsection{Residual transformations of the 1st kind and transformations of the 2nd kind}\label{Residual transformations of the 1st kind and transformations of the 2nd kind}

If $\H$ is a subgroup of the full gauge group, the dressed fields \eqref{dressed-fields} display so-called \emph{residual transformations of the 1st kind} under what remains of the gauge group after reduction via the DFM. 
Suppose, in fact, that one is given a $\K$-dressing field, $u^\kappa=\kappa^{-1}u$, $\kappa \in \K$, with $\K \subseteq \H$ the equivariance group of $u \in \mathcal{D}r[G,\K]$, with $G$ the target group ($G$ s.t. either $H\supseteq G \supseteq K$ or $G\supseteq H$, with $K$ and $H$ rigid groups -- $H$ structure group in bundle differential geometric terms).
In this case, dressed fields are expected to display residual (gauge) transformations of the 1st kind under what remains of the gauge group (not fully reduced).
If $K \triangleleft H$ (i.e., $K$ is a normal subgroup of $H$), then $H/K=:J$ is a Lie group, with associated gauge subgroup $\J = \H/\K$, $\K \triangleleft \H$, and one is left with well-defined $\J$-gauge residual transformations of the dressed fields.
For instance, if $u$ is a $\K$-dressing field s.t. $u^\eta = \eta^{-1} u \eta$, for $\eta \in \J$, then dressed fields are $\J$-gauge variables (whose $\J$-gauge transformations can be simply obtained by taking their $\H$-gauge transformations and formally replacing $\gamma \in \H \rightarrow \eta \in \J$).
However, in concrete applications of the DFM and in order to fully exploit its relational interpretative potential, what one should do is completely reduce the gauge group $\H$.

In the DFM one also typically encounters possible ``ambiguities'' in the choice, or construction, of dressing field. 
Indeed, any two dressing fields $u=u[\upphi]$ and $u'=u'[\upphi]$, both satisfying the defining property of a dressing field, are a priori related as $u'=u\xi$ for some $\H$-invariant map $\xi:M \rarrow H$. 
Under pointwise multiplication, such maps form 
the group of so-called \emph{transformations of the 2nd kind}, $\G\defeq\{\xi:M \rarrow H\, |\, \xi^\gamma=\xi\}$.\footnote{In the literature we developed previously on this topic, we had referred to them as \emph{residual} transformations of the 2nd kind. However, we later renamed them simply transformations of the 2nd kind, removing the adjective ``residual'' in order to avoid confusion with residual gauge symmetries. Indeed, they are not a subgroup (or a remnant) of the gauge symmetry that has been reduced through the dressing procedure.}
Since dressing fields depend on $\upphi$, elements of $\G$ may also be $\upphi$-dependent, i.e., $\xi=\xi[\upphi]$. 
Writing the action of $\G$ on a dressing field as $u^\xi=u\xi$, 
the $\G$-transformations of dressed fields $\upphi^u$ are a priori: 
\begin{equation}
\label{2nd-kind-trsf}
    (\upphi^u)^\xi \defeq (\upphi^\xi)^{u^\xi}=(\upphi^\xi)^{u\xi}, 
    \quad \xi\in \G,
\end{equation}
whose result depends on the $\G$-transformation of the bare fields $\upphi$, if any.
Transformations of the 2nd kind have been shown to encode \emph{physical reference frame covariance} \cite{JTF-Ravera2024NRrelQM}. 
In this case, the group $\G$ is ``discrete'' in the sense that it enumerates, among the field-theoretic d.o.f., the available options that may serve as a physical reference frame, i.e., as a dressing field.

\subsection{Dressing Field Method \emph{vs} gauge fixing}\label{Dressing Field Method vs gauge fixing}

Let us highlight some differences between the DFM and gauge fixing, following, e.g., \cite{JTF-Ravera2024gRGFT,JTF-Ravera2024-SUSY,JTF-Ravera2025DFMSusyReview&Misu,Berghofer-Francois2024}.
It is evident, by the very definition of a dressing field, that is from the way it gauge-transforms, that $u \notin \H$. 
This is a crucial fact of the DFM: Despite the formal analogy with gauge-transformed fields, the dressed fields are not gauge transformations.
The dressed fields $\{A^u, \vphi^u\}$ in \eqref{dressed-fields} are not a point in the gauge $\H$-orbit $\O^\H_{\{A, \vphi\}}$ of $\{A, \vphi\}$. Hence, $\{A^u, \vphi^u\}$ must not be confused with a gauge fixing of the bare variables $\{A, \vphi\}$, namely with a point on a gauge-fixing slice. 
Contrary to a gauge fixing, the ``dressing operation''  is not a map from $\Phi$ (field space) to itself, but from $\Phi$ to the space of dressed fields $\Phi^u$, only isomorphic to a sub-bundle of $\Phi$. 
In particular, if one considers a complete symmetry reduction via an $\H$-dressing field, then $\Phi^u$ is readily understood as a \emph{coordinatization of the moduli space} $\M$ -- or of a region $\U \subset \M$.
Furthermore, dressed fields are \emph{relational variables}, while representative fields located on a gauge-fixing slice are not. 

For a better understanding of the difference between gauge fixing and the dressing via the DFM, let us start by shortly discussing in the following the features of gauge fixing.

\subsubsection{Gauge fixing}\label{Gauge fixing}

\begin{figure}[ht]
\begin{center}
\includegraphics[width=0.7\textwidth]{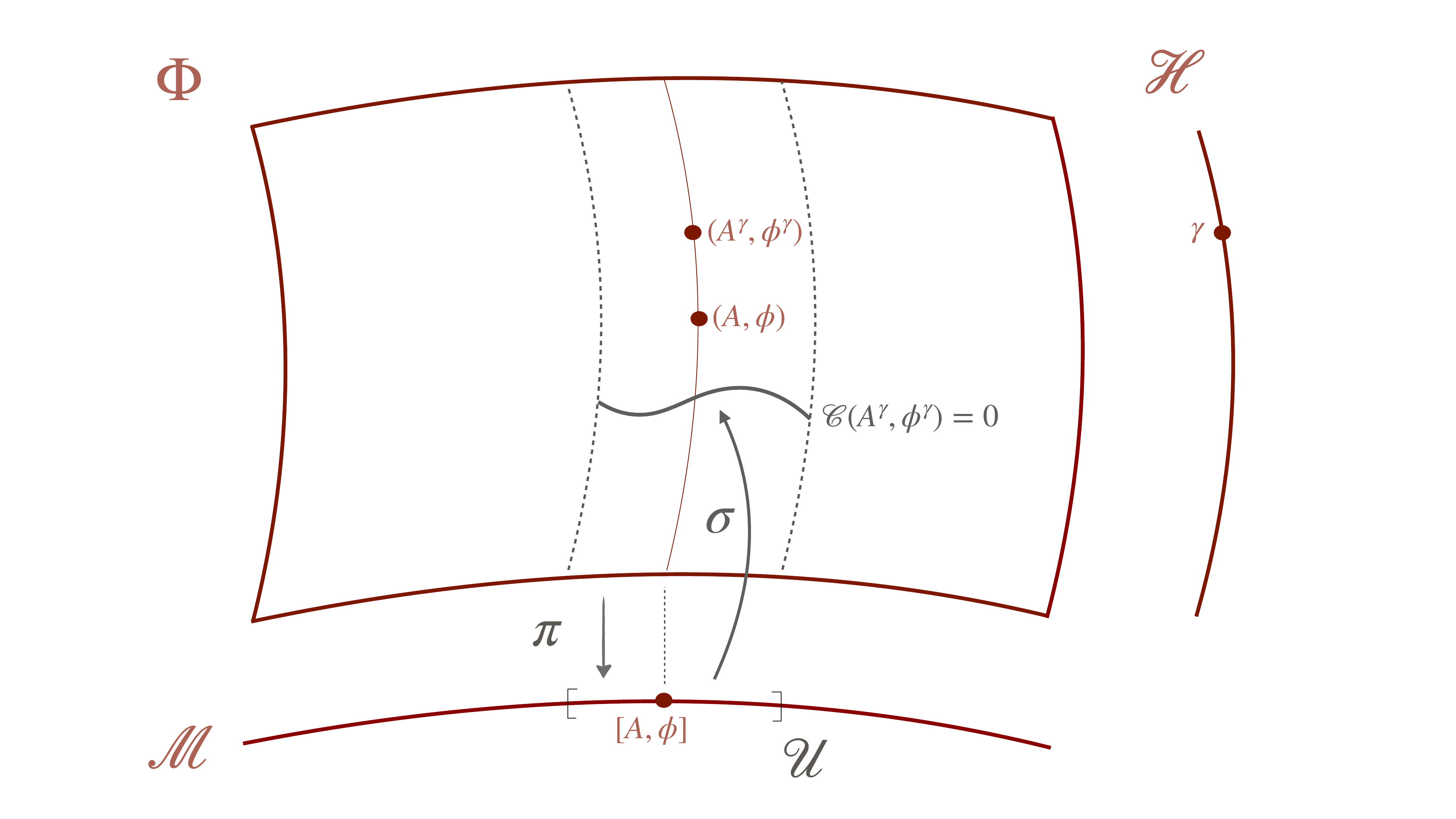}
\caption{Gauge fixing in $\Phi$, i.e., a choice of local section $\sigma$ of $\Phi$. The gauge-fixing slice is the image of $\sigma$. Picture extracted from \cite{JTF-Ravera2025DFMSusyReview&Misu}.}
\label{figure:bundlegeomfs}
\end{center}
\end{figure}

In gRGFT, it is often considered convenient to somehow restrict to those variables $\{A, \vphi\}$ satisfying particular functional properties. This can make computations more manageable and is typically considered a necessary step towards quantization.
When restrictions are imposed by exploiting the gauge freedom of the fields, they are referred to as ``gauge-fixing'' conditions.  
The functional restrictions define a ``slice'' in $\Phi$, namely cutting across gauge orbits once. 
This selects a single representative in each. In other words, as schematically represented in Figure \ref{figure:bundlegeomfs}, a gauge fixing is a \emph{choice of local section} of the field space $\Phi$ seen as a bundle, namely $\s: \U\subset\M \rarrow \Phi$.

Concretely, a gauge fixing is specified by a condition taking the form of an algebraic and/or differential equation on fields, implemented by using the gauge freedom: $\C(A^\gamma, \vphi^\gamma)=0$. 
Such a section does not exist globally, unless $\Phi$ is a trivial bundle, i.e., $\Phi=\M \times \H$ (cf. the literature on the Gribov-Singer obstruction/ambiguity, starting with \cite{Singer1978, Singer1981, Fuchs1995}).

\subsubsection{Difference between dressing via the DFM and gauge fixing}\label{Difference between dressing via the DFM and gauge fixing}

\begin{figure}[ht]
\begin{center}
\includegraphics[width=0.7\textwidth]{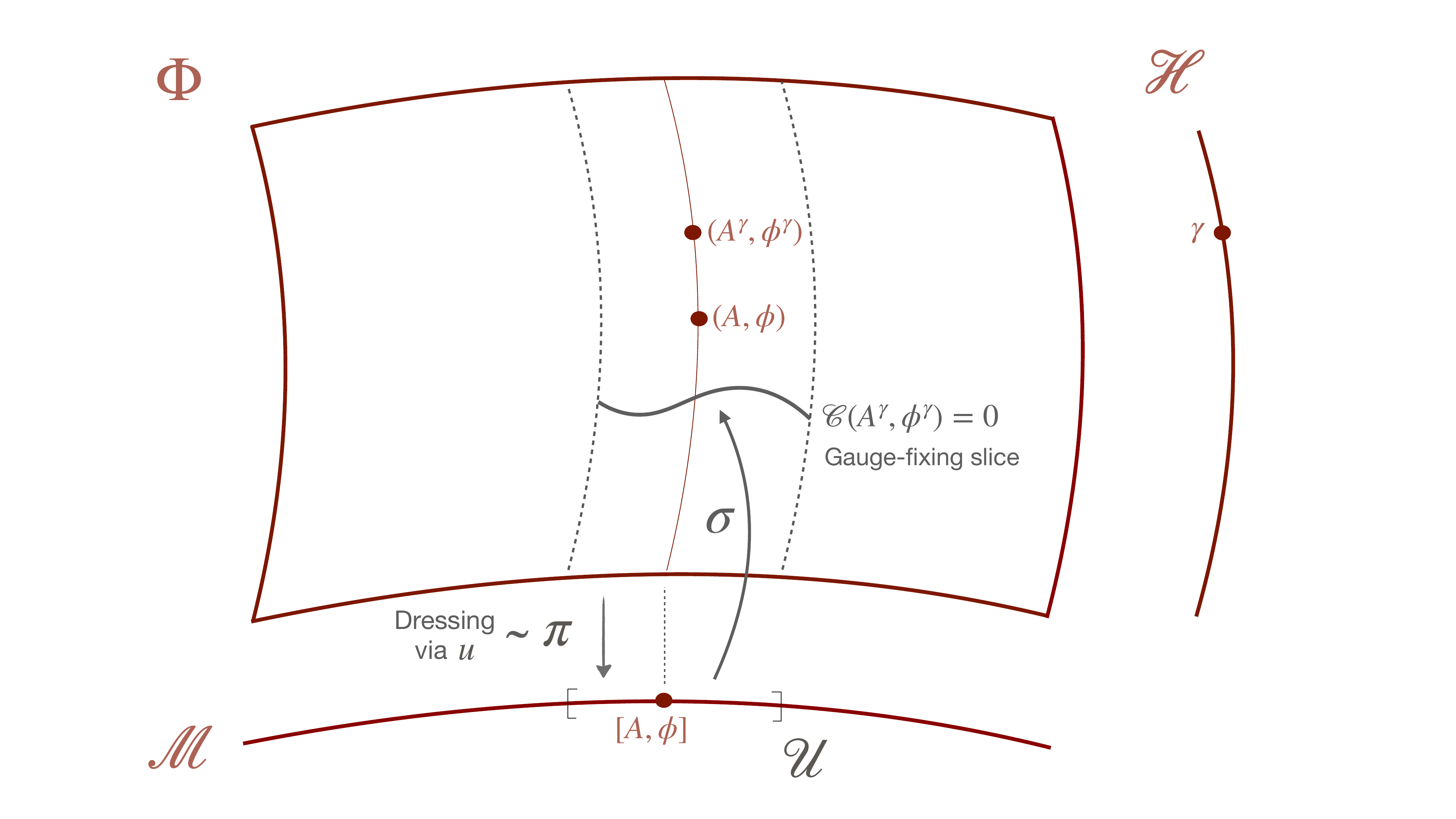}
\caption{Difference between dressing as done via the DFM and gauge fixing in field space $\Phi$. Picture extracted from \cite{JTF-Ravera2025DFMSusyReview&Misu}.}
\label{figure:DFMvsgaugefixing}
\end{center}
\end{figure}

As we already mentioned, comparing the definition of the gauge group and that of a dressing field, it is evident that $u \notin \K$. 
It follows that a $\upphi$-dependent dressing field $u=u[\upphi]$, which by definition transforms as $u[\upphi]^\kappa:=u[\upphi^\kappa]=\kappa\- u[\upphi]$, 
cannot be misconstrued as a $\upphi$-dependent gauge group element $\gamma[\upphi]$, which by definition is s.t. $\gamma[\upphi]^\kappa:=\gamma[\upphi^\kappa]=\kappa\- \gamma[\upphi] \kappa$.
This is a crucial fact of the DFM: Despite the formal analogy, the dressed fields \eqref{dressed-fields} are not gauge transformations.
The dressed fields $\{A^u, \vphi^u\}$ are \emph{not} a point in the gauge $\K$-orbit $\O^\K_{\{A, \vphi\}} \subset \O^\H_{\{A, \vphi\}}$ of $\{A, \vphi\}$. 
Hence, $\{A^u, \vphi^u\}$ must not be confused with a gauge fixing of the bare variables $\{A, \vphi\}$, namely with a point on a gauge-fixing slice ($\S$). 
Contrary to a gauge-fixing $\Phi \rarrow \S \subset \Phi$, the ``dressing operation''  is not a map from $\Phi$ to itself, but from $\Phi$ to the space of dressed fields $\Phi^u$.
%, only isomorphic to a subbundle of $\Phi$. 
In particular, if one considers a complete symmetry reduction via an $\H$-dressing field, then $\Phi^u$ is readily understood as a \emph{coordinatization of the moduli space} $\M$ -- or of a region $\U \subset \M$ over which the field-dependent dressing field $u: \Phi_{|\U} \rarrow \D r[G, \H]$ is defined: the one-to-one mapping $(\Phi_{|\U})^u \leftrightarrow \U \subset \M$ may indeed be seen as a ``coordinate chart''. 
It follows that performing the dressing procedure allows to work with the physical d.o.f. which are not accessible in any way to direct computations through gauge fixing. 
Furthermore, as highlighted above, dressed fields $\upphi^u=\{A^u, \vphi^u\}$ are invariant, relational variables, while representatives fields located on a gauge-fixing slice $\S$ are neither of the two things. 
This distinction applies to gauge fixings, however they are implemented.
The difference between the implementation of a (full) dressing operation via the DFM and a gauge fixing in the field space bundle $\Phi$ is schematically represented in Figure \ref{figure:DFMvsgaugefixing}. 

Finally, we remark that the theory
confronted to experimental tests is always the dressed one, not a gauge-fixed version of it, see \cite{JTF-Ravera2024gRGFT} and \cite{Berghofer-Francois2024,Masson-et-al2024} for further detailed discussions of the technical and conceptual distinction between gauge fixing and dressing.

\subsection{Examples of applications of the DFM to gauge symmetries}\label{Examples of applications of the DFM to gauge symmetries}

Here we provide some examples of applications of the DFM to construct variables that are automatically invariant and manifestly relational in GFT. 
We will also consider cases in which a dressing field can be extracted from functional constraints on the available field-theoretic variables, typically imposed in gauge-fixing procedures. 
We will see how, within the DFM, these constraints instead become identities that are automatically satisfied by the dressed fields.

\subsubsection{Three-dimensional non-Abelian Chern-Simons theory}\label{Three-dimensional non-Abelian Chern-Simons theory}

As a ``toy model'' to start with, we consider the case of the three-dimensional non-Abelian CS theory, whose field content is given by $\upphi=\lbrace{ A \rbrace}$, where $A$ is an $\mathfrak{su}(n)$-valued 1-form field. 
The gauge group is $\H=\SO(n)$, and $A$ gauge-transforms as
\begin{align}
    A^\gamma := \gamma^{-1} A \gamma + \gamma^{-1} d \gamma , \quad \gamma \in \H=\SO(n).
\end{align}
The curvature, defined as $F:=dA + \tfrac{1}{2}[A,A]$, gauge-transforms as
\begin{align}
    F^\gamma := \gamma^{-1} F \gamma .
\end{align}
The Lagrangian of the theory is
\begin{align}
    L_{\text{CS}}(A) = \Tr \left(AdA + \tfrac{2}{3}A^3 \right) .
\end{align}
Under gauge transformations, the Lagrangian transforms as
\begin{align}
    L_{\text{CS}}(A)^\gamma = L_{\text{CS}}(A^\gamma) =  L_{\text{CS}}(A) + \Tr \left( d(\gamma d \gamma^{-1}A) - \tfrac{1}{3} (\gamma^{-1}d\gamma)^3\right) ,
\end{align}
where we can observe that $\Tr \left( d(\gamma d \gamma^{-1}A) - \tfrac{1}{3} (\gamma^{-1}d\gamma)^3\right)=: c(A,\gamma)$ is a group cocycle (this is manifest in the differential-geometric framework of cocyclic geometry). 
Note that, at the finite level, the CS Lagrangian is not even quasi-invariant, since it does not transform merely by a boundary term (as instead happens when the theory is treated at the level of infinitesimal transformations).
The field equation of the theory, obtained by varying the Lagrangian w.r.t. $A$, is $F=0$.

We now turn to the application of the DFM. If a dressing field $u$, s.t. $u^\gamma = \gamma^{-1}u$, for $\gamma \in \H=\SO(n)$, is extracted/constructed (as is typically done using the holonomies of the connection, or can be achieved \emph{ad hoc}, for instance through the inclusion of so-called \emph{edge modes}\footnote{Which, indeed, are typically an instance of \emph{ad hoc} dressing fields, namely dressing fields imported from outside into the theory -- not built from the original field content.}), then one can construct the dressed field $A^u$ and the dressed curvature $F^u$ (using the DFM rule of thumb):
\begin{align}
    A^u := u^{-1} A u + u^{-1} d u , \quad F^u := u^{-1} F u,
\end{align}
which are s.t.
\begin{align}
    (A^u)^\gamma = A^u , \quad (F^u)^\gamma = F^u ,
\end{align}
hence gauge-invariant.
The dressed Lagrangian is then
\begin{align}
    L_{\text{CS}}(A)^u = L_{\text{CS}}(A^u) =  L_{\text{CS}}(A) + \Tr \left( d(u d u^{-1}A) - \tfrac{1}{3} (u^{-1}du)^3\right) ,
\end{align}
and it is \emph{strictly} gauge-invariant by construction (it is a functional of the dressed field $A^u$). The dressed field equation reads $F^u=0$, with $F^u := u^{-1} F u$.

\subsubsection{Maxwell electromagnetism and the ``Lorenz gauge as a dressing'' via the DFM}\label{Maxwell electromagnetism and the ``Lorenz gauge as a dressing'' via the DFM}

Let us now consider the case of pure Maxwell electromagnetism (EM), with Abelian gauge group $\H=\U(1)$, whose field content is given by $\upphi = \lbrace{A \rbrace}$, where $A = A_\mu dx^{\,\mu}$ is the 1-form gauge potential. The field $A$ transforms under gauge symmetry as
\begin{align}
    A^\lambda = A + d\lambda,
\end{align}
or, in components (with $\mu,\nu=0,1,\ldots,4$),
\begin{align}
    A_\mu^\lambda = A_\mu + \partial_\mu \lambda .
\end{align}
We note that $\lambda^{\lambda'} = \lambda$, for $\lambda,\lambda' \in \H$, since conjugation is trivial in the Abelian case.
The field strength is $F_{\mu \nu}:=\partial_{[\mu}A_{\nu]}$, it fulfills the Bianchi identities $\partial_{[\mu}F_{\nu \rho]}=0$, and the Lagrangian of the theory reads
\begin{align}
\label{lagrEM}
    L_{\text{EM}} = - \tfrac{1}{4} F_{\mu \nu} F^{\,\mu \nu} .
\end{align}
Note that $F_{\mu \nu}$ is gauge-invariant, and so is the Lagrangian \eqref{lagrEM}, which is entirely written in terms of the field strength. 

In the following, we will focus only on the kinematics, since our aim is to show that the covariant functional constraint commonly referred to as the \emph{Lorenz gauge} in EM, when considered on the dressed field variable, can in fact be used to extract a dressing field for the $\U(1)$ symmetry.

Let us first consider the Lorenz gauge 
\begin{align}
\label{Lorenzgauge}
    \partial^{\,\mu} A_\mu=0
\end{align}
and perform a standard ``gauge fixing'' ($\C(A)=\partial^{\,\mu} A_\mu=0$).
For consistency of the ``gauge-fixing constraint'', we shall require the gauge-transformed field to still fulfill the same constraint ($\C(A^\lambda)=\partial^{\,\mu} A_\mu^
\lambda=0$), that is
\begin{align}
    \partial^{\,\mu} A_\mu^\lambda = 0 \quad \Rightarrow \quad \Box \lambda = 0, 
\end{align}
with $\Box:=\partial^{\,\mu}\partial_\mu$.
This constrains the gauge parameter $\lambda$ and encodes what are commonly referred to as \emph{residual gauge transformations}.\footnote{Note that the Lorenz gauge in the Maxwell theory does not affect/reduce the propagating d.o.f., but constrains the field space.}
In general, an ``ideal'' (or ``perfect'') gauge fixing (locally, if globally not possible due to Gribov ambiguities) is specified by a condition (constraint) that selects one and only one representative by gauge orbit: 
a ``thin'' slice in field space, in the sense that no subgroup of the gauge group is an automorphism of the slice.
A partial (on ``non-ideal'', ``imperfect'') gauge fixing is a condition $C(\upphi)$ that selects a subspace of field space on which a subgroup of the gauge group acts as an automorphism. 
This group is the group of ``residual (unfixed) gauge symmetries'':
that subgroup is specified by condition(s) $C'(\lambda)$ on the gauge parameter $\lambda \in \mathcal{H}$,
compatible with its definition as a gauge group element (e.g., in the $\U(1)$ case, being gauge-invariant).

One way to apply the DFM in the case of Maxwell EM (see, e.g., \cite{Berghofer-Francois2024}), consists in starting by considering the ``Lorenz functional constraint''
\begin{align}
    \partial^{\,\mu} A^u_\mu = 0 
\end{align}
on the field variable
\begin{align}
\label{dressedAmu}
    A^u_\mu := A_\mu + \partial_\mu u ,
\end{align}
and solve it explicitly for $u$:
\begin{equation}
    \partial^{\,\mu} A^u_\mu = \partial^{\,\mu} (A_\mu + \partial_\mu u) = 0 \quad \Rightarrow \quad u=u[A] = - \Box^{-1} (\partial^{\,\mu} A_\mu ) + g = u_0 + g , \quad \text{s.t.} \quad \Box g = 0 .
\end{equation}
We then check the gauge transformation of $u=u[A]$ (and $g$) to see if we are dealing with a gauge group element or a dressing field. 
First of all, we observe that $u_0=u_0[A]$ is automatically a (field-dependent) dressing field, as
\begin{align}
    u_0^\lambda = u_0 - \lambda ,
\end{align}
as expected for a dressing field in this Abelian model. 
Hence $u_0 \notin \H$.
On the other hand, a gauge group element, for the model at hand, is invariant:
\begin{align}
    \lambda^{\eta} = \lambda, \quad \lambda, \eta \in \H .
\end{align}
As far as $g$ is concerned, the natural gauge transformation for it would yield $g^\lambda=g$ (``null hypothesis''). 
On the other hand, one may also, in principle, regard it as an object transforming differently. 
In particular, we may consider $g^\lambda = g+\lambda$, which would make $u$ an invariant object itself.
Thus, at this stage, we can say that two main scenarios arise:
\begin{itemize}
    \item[A)] The case $g^\lambda = g+\lambda$, that is $u^\lambda=u$. This is a ``constructive way'' of implementing a (\emph{field-dependent}) gauge fixing. 
    In fact, $u$ can be now thought of to as a \emph{field-dependent gauge parameter}.
    Note that $g^\lambda = g+\lambda$, together with $\Box g=0$ (by definition), implies, for consistency (``to preserve the space of $g$''), $\Box \lambda=0$. 
    The latter encodes the \emph{residual gauge symmetry}.
    Thus, we now have
    \begin{align}
        A^u_\mu = A^{u_0}_\mu + \partial_\mu g , \quad \text{s.t.} \quad \partial^\mu A^u_\mu \equiv 0 \quad \text{by construction},
    \end{align}
    and $A^u$ is \emph{not} gauge-invariant (notice that, for consistency, $\Box \lambda=0$).  
    In fact, $A^u$ gauge transforms as
    \begin{align}
    \label{AuTr-gf}
        (A^u_\mu)^\lambda  = (A^{u_0}_\mu)^\lambda + \partial_\mu g = A^{u_0}_\mu + \partial_\mu g + \partial_\mu \lambda  = A^u_\mu  + \partial_\mu \lambda .
    \end{align}
    Here it is important to stress that the role of $g$ is just to have that $A^u_\mu$ is still a gauge potential, that is to support the correct transformation still.
    It is declared by hand to transform in such a way to produce, essentially, an \emph{ad hoc}, field-independent, ``inverse of a dressing field''.
    This becomes clearer in group-theoretic terms. In fact, we are in the following situation. We built, more formally,
    \begin{align}
        U\equiv U_0 \tilde{U}^{-1},
    \end{align}
    which gauge-transforms as
    \begin{align}
        U^\gamma = (U_0)^\gamma (\tilde{U}^\gamma)^{-1} = \gamma^{-1} U_0 \tilde{U}^{-1} \gamma = \gamma^{-1} U \gamma ,
    \end{align}
    that is, $U$ is a \emph{field-dependent gauge parameter} (field-dependent element of $\H$).
    In the case at hand, in particular, we have
    \begin{equation}
    \begin{aligned}
        & \tilde{U}= e^{-g}, \quad \text{s.t.} \quad \tilde{U}^\gamma = e^{-\lambda} e^{-g} = e^{-(g+\lambda)} = e^{{-g}^\lambda}, \quad \text{and} \quad U_0 = e^{u_0},  \\
        & U= e
        ^U = U_0[A] \tilde{U}^{-1}  = e^{u_0[A]} e^g = e^{u_0[A] +g},
    \end{aligned}
    \end{equation}
    with $g$ being the ad hoc, field-independent object introduced to perform this constructive gauge fixing.
    
    \item[B)] The case $g^\lambda=g$, and $u^\lambda = u_0^\lambda + g^\lambda = u_0 - \lambda + g = u - \lambda$. Hence, $u$ is a \emph{dressing field}, and we are now fully in the realm of the DFM, with $g$ encoding transformations of the 2nd kind.
\end{itemize}
The fact that one can formally solve a functional constraint on the fields in order to extract a dressing field is now well-known in the literature (see, for example, \cite{Berghofer-Francois2024,JTF-Ravera2024-SUSY,JTF-Ravera2024ususyDFM,JTF-Ravera2025DFMSusyReview&Misu}). 
Other ``gauge fixings'' can thus be reformulated and reinterpreted within the DFM framework (this is the case, for instance, for the axial gauge, the harmonic gauge, the de Donder gauge, the unitary gauge, etc.).\footnote{For instance, more explicitly, the  ``\emph{axial gauge}'' for a vector gauge potential $A_{\mu}$ in QED reads $n^{\,\mu}A_{\mu}=0$, 
with $n^{\,\mu}$ a constant 4-vector -- cf., e.g., \cite{Leibbrandt-Richardson1992}. 
In fact, by solving explicitly the axial constraint, one ends up building  a (non-local) dressing field $u[A, n]$; the latter then yields the $\U(1)$-invariant (self-)dressed field $A^u_\mu\defeq A_\mu+ \d_\mu u[A, n]$.}  
We note that, in the case of EM and the Lorenz gauge just analyzed, the variable $A^u$ provides an example of \emph{self-dressing}, and the extracted dressing field is \emph{non-local}, indicating that the reduced symmetry is \emph{substantive} for Maxwell theory \cite{Francois2018}: the $\U(1)$ gauge symmetry is reduced, building manifestly gauge-invariant variables, at the ``price'' of locality.
We further observe that, whenever a dressing field is extracted by solving a constraint defined in terms of an operator acting on field-theoretic variables (as in the EM case above), one should always expect an ambiguity in the dressing: 
It arises from the fact that the dressing field is defined only up to elements in the kernel of such an operator, and can be redefined by means of them (this is encoded in transformations of the 2nd kind).

\subsubsection{The Abelian Higgs model without SSB}\label{The Abelian Higgs model without SSB}

The DFM offers a reinterpretation of the Brout–Englert–Higgs (or Higgs, for short) mechanism, in terms of dressed fields, that avoids the need to postulate the SSB.
As an example, let us therefore consider the simple Abelian Higgs model (see, e.g., \cite{Francois2018} for the DFM treatment of this model),
whose kinematics is provided by the field content $\upphi=\lbrace{A,\varphi \rbrace}$, with $\varphi \in \Omega^0(M,\mathbb{C})$, or $\varphi \in C^\infty (M,\mathbb{C})$, a complex scalar, 
which gauge-transform under the $\U(1)$ gauge symmetry of the model as
\begin{align}
     A^\gamma = \gamma^{-1} A \gamma + \gamma^{-1} d \gamma , \quad \varphi^\gamma = \gamma^{-1} \varphi, \quad \gamma \in \H=\U(1).
\end{align}
Note that in the Abelian $\U(1)$ case the gauge transformation of $A$ simply yields $A+d\lambda$, as $\gamma=e^\lambda$. 
We also define the covariant derivative $D\vphi:=d\vphi + A \vphi$ (formally, let us omit $i$ factors and the coupling constant, for simplicity) and the field strength $F=dA$ -- in components, $F_{\mu \nu}=\partial_{[\mu}A_{\nu]}$.

Let us now summarize the ``standard'' steps of the Higgs mechanism and, especially, the interpretation with SSB. The narrative goes as follows:
\begin{enumerate}
    \item The dynamics of the Abelian Higgs model is given by the Lagrangian
    \begin{align}
    \label{lagrabhi}
        L = F \star F + D\vphi^* \star D \vphi + \alpha \vphi^* \vphi + \beta (\vphi^* \vphi), \quad \beta >0 ,
    \end{align}
    where $\vphi^*$ is the complex conjugate of $\vphi$ and $\star$ denotes the Hodge star operator. The potential is $V(\vphi)=\alpha \vphi^* \vphi + \beta (\vphi^* \vphi)$, where $\alpha$ and $\beta>0$ are constants ($\alpha$ yields the mass in the scalar EM model). 
    In components, the Lagrangian \eqref{lagrabhi} reads\footnote{In the following, we will omit the index notation, which can be easily derived and found in standard textbooks on field theory.}
    \begin{align}
        L = - \tfrac{1}{4} F_{\mu \nu} F^{\,\mu \nu} - D_\mu \vphi^* D^{\,\mu} \vphi + \alpha \vphi^* \vphi + \beta (\vphi^* \vphi) .
    \end{align}
    \item The vacuum is studied, where the functional $\tfrac{\delta V}{\delta \vphi^*}$ evaluated at $\vphi_0$ should be $0$ by definition, that is
    \begin{align}
    \label{varpot}
        \tfrac{\delta V}{\delta \vphi^*} (\vphi_0) = [\alpha \vphi + \beta \vphi (\vphi^* \vphi)]_{\vphi_0} = [\vphi (\alpha + 2 \beta (\vphi^* \vphi))]|_{\vphi_0} = 0.
    \end{align}
    \item The solutions of \eqref{varpot} are analyzed. We find
    \begin{align}
        \vphi_0 = 0 ,
    \end{align}
    which is $\U(1)$-invariant, and
    \begin{align}
    \label{degsol}
        |\vphi_0|^2 = - \tfrac{\alpha}{2\beta} \quad \Rightarrow \quad \vphi_0 = \sqrt{-\tfrac{\alpha}{2 \beta}} =: v ,
    \end{align}
    where we denote by $v$ the vacuum expectation value (v.e.v.) of $\vphi$.
    Using \eqref{degsol}, we may then write
    \begin{align}
        \vphi_0 = v e^{i \theta} ,
    \end{align}
    that is we write $\vphi_0$ in terms of the v.e.v. $v$ and a phase factor $e^{i \theta}$.
    \item We observe that for $\alpha>0$ we have a unique vacuum, while for $\alpha<0$ we have a degeneracy of vacua (the manifold of vacua is a $\U(1)$-orbit; each point is not invariant).
    Hence, a SSB of the $\U(1)$ symmetry is said to occur -- and, for each broken generator (one in the $\U(1)$ case), there exists a Goldstone boson ``eaten'' to give mass to the gauge field).
    It is precisely here that, as we will see shortly, the DFM provides an alternative interpretation without invoking SSB (a possibility already considered, e.g., by Englert and Higgs themselves), from which the standard application of the Higgs mechanism then follows. 
    For completeness, we therefore briefly review the latter below (cf. steps 5. and 6.).
    \item The expansion around the vacuum is then considered, that is
    \begin{align}
    \label{expvac}
        \vphi = \vphi_0 + H ,
    \end{align}
    where $H$ is the Higgs field/boson.
    By replacing \eqref{expvac} in the covariant derivative, we get (note that $d\vphi_0=0$, as $\vphi_0$ is constant)
    \begin{align}
        D \vphi = d \vphi + A \vphi = d (\vphi_0 + H) + A (\vphi_0 + H) = D H + A \vphi_0 .
    \end{align}
    On the space of vacuum solutions we have
    \begin{align}
        \vphi_0 = 0 \quad \Rightarrow \quad D \vphi = D H ,
    \end{align}
    and 
    \begin{align}
        \vphi_0 = \sqrt{-\tfrac{\alpha}{2 \beta}} =: v  \quad \Rightarrow \quad D \vphi = D H + A v e^{i \theta}.
    \end{align}
    Considering the latter together with the ``unitary gauge'' $\theta=0$, we are left with
    \begin{align}
    \label{substres}
        D \vphi = D H + A v .
    \end{align}
    Plugging this result back into the Lagrangian, we may then analyze the result term by term. 
    In particular, the kinetic term for the scalar boils down to (we use $A^*=-A$)
    \begin{align}
         D\vphi^* \star D \vphi = DH \star DH - v^2 A^2 ,
    \end{align}
    where we recognize the kinetic term for the Higgs field and the mass term for $A$.
    Plugging \eqref{substres} back into the potential $V(\vphi)$, one then also gets higher terms, yielding the mass and self-interaction terms for the Higgs field.
\end{enumerate}
We now revisit steps 1. through 4. under the lens of the DFM, and show how this framework avoids invoking SSB.
Kinematically, let us thus consider the polar decomposition of $\vphi$:
\begin{align}
    \label{polardec}
    \vphi = \rho e^{i \theta} = \rho u ,
\end{align}
where we can in fact check that the phase factor gauge-transforms, under the $\U(1)$ gauge symmetry, as a dressing field,
\begin{align}
    u^\gamma = \gamma^{-1} u , \quad \gamma \in \U(1) .
\end{align}
Therefore, we can build the dressed fields
\begin{align}
    A^u = A + u^{-1}du, \quad \vphi^u = \rho ,
\end{align}
and $F^u=F$ (in the simple Abelian case).
The dressed fields are $\U(1)$-invariant variables.
We may now retrace steps 1. through 4. in terms of the dressed variables.
\begin{enumerate}
    \item The dressed Lagrangian is
    \begin{align}
        L^u = L(A^u,\rho) = F \star F + D^u \rho^* \star D^u \rho + \alpha \rho^2 + \beta \rho^4 ,
    \end{align}
    where $D^u \rho = d \rho + A^u \rho = d \rho + A \rho + u^{-1}du \rho$.
    Note that the potential is $V=\alpha \rho^2 + \beta \rho^4$, with $\beta>0$.
    \item Studying the vacuum, we get
    \begin{align}
        \tfrac{\delta V}{\delta \rho} (\rho_0) = (2 \alpha \rho + 4 \beta \rho^3)|_{\rho_0} = [2 \rho (\alpha + 2 \beta \rho^2)]|_{\rho_0} = 0.
    \end{align}
    \item The solutions now read
    \begin{align}
        \rho_0 = 0 
    \end{align}
    and
    \begin{align}
        \rho_0^2 = - \tfrac{\alpha}{2 \beta} \quad \Rightarrow \quad \rho_0 = \sqrt{-\tfrac{\alpha}{2\beta}} =: v .
    \end{align}
    But $\rho$ is a $\mathbb{R}^+$-valued function. 
    \item So, for both $\alpha>0$ and $\alpha<0$, that is, in both phases, the vacuum is now unique.
    The $\U(1)$ gauge symmetry is ``reduced'' in both phases (note that this occurs already at the kinematical levels, even before considering $\alpha>0$ and $\alpha<0$), and there is no need to invoke SSB.
\end{enumerate}
Steps 5. and 6. then follow analogously to what was seen previously, but without the interpretation in terms of SSB, and expressed in terms of the dressed variables. 
In fact, we may observe that there is no causal relation between SSB and the notion of ``mass acquisition'' via the Higgs mechanism.

\subsubsection{Lorentz dressing in GR}\label{Lorentz dressing in GR}

Another example of dressing via the DFM is provided by the ``Lorentz dressing'' in GR (and, more generally, in general-relativistic theories). 
This procedure is well-known in the literature and is often performed to pass from the language of differential forms (vielbein and spin connection) to the metric formulation in components (in terms of the metric and Christoffel symbols), although it is not often identified explicitly as a dressing. 
In fact, what takes place in this procedure is a reduction of the local Lorentz symmetry.
Let us briefly review it.

Let us consider an Einstein-Cartan theory in four  dimensions (also in the presence of a non-vanishing cosmological constant). 
Here we focus on the kinematics, and refer the interested reader to \cite{Francois2021} for the full development of the dynamics, which follows straightforwardly once the kinematics has been analyzed.
We consider a Cartan connection $\bar{A} = A + e$, where $A$ is $\mathfrak{so}(1,3)$-valued and $e$ is the soldering form, which is $\mathbb{R}^4$-valued. 
The gauge group is the Lorentz group $\SO(1,3)$, and it acts on the local representatives of the Cartan connection as
\begin{align}
A^\gamma = \gamma^{-1} A \gamma + \gamma^{-1} d\gamma,
\quad
e^\gamma = \gamma^{-1} e, \quad \gamma \in \SO(1,3).    
\end{align}
The full Cartan connection transforms as $\bar{A}^\gamma = \gamma^{-1} \bar{A} \gamma + \gamma^{-1} d\gamma$, for $\gamma \in \SO(1,3)$.
It follows that an $\mathrm{SO}(1,3)$-dressing field can be extracted from the soldering form (vielbein) itself, that is, from a component of the Cartan connection.
Indeed, given a coordinate system $\lbrace{ x^{\,\mu}\rbrace}$ on $U \subset M$, the soldering is $e=e^a_{\,\mu}dx^{\,\mu}$,
so the map ${\bf{e}}:= e^a_{\,\mu}:U \to GL(4)$ is s.t. ${\bf{e}}^\gamma=\gamma^{-1}{\bf{e}}$.
Therefore, we have a field-dependent (\emph{local}) dressing field ${\bf{u}}(\bar{A})={\bf{e}}$, s.t.
\begin{align}
    {\bf{u}}[\bar{A}]^\gamma = {\bf{u}}[\bar{A}^\gamma] = {\bf{e}}^\gamma = \gamma^{-1} {\bf{e}} = \gamma^{-1} {\bf{u}}[\bar{A}].
\end{align}
Said otherwise, the tetrad field is a Lorentz dressing field.
We can therefore write the $\SO(1,3)$-invariant, dressed Cartan connection:
\begin{align}
    \bar{A}^{\bf{u}} = {\bf{u}}^{-1} \bar{A} {\bf{u}} + {\bf{u}}^{-1} d {\bf{u}} =: \bar{\Gamma} ,
\end{align}
whose components are
\begin{equation}
\begin{aligned}
    A^{\bf{u}} & = {\bf{e}}^{-1} A {\bf{e}} + {\bf{e}}^{-1} d {\bf{e}} =: \Gamma , \\
    e^{\bf{u}} & = {\bf{e}}^{-1} e = dx ,
\end{aligned}
\end{equation}
where $dx = {\delta^{\,\mu}}_\rho dx^\rho$ and
$\Gamma = {\Gamma^{\,\mu}}_{\lambda} ={\Gamma^{\,\mu}}_{\lambda\,,\,\rho} \,dx^{\,\rho} $ is the familiar linear connection -- which has values in $\text{Lie}GL(4)$.
One can then also dress the Cartan curvature, whose components are the Lorentz curvature and the torsion 2-form, as well as the Lagrangian of the theory, thereby obtaining the metric formulation of GR in the presence of a cosmological constant (see, e.g., \cite{Francois2021}).
We can observe that, in the case of the tetrad field understood as a Lorentz dressing, we have (the Jacobian of) coordinate changes acting on the right (these being a case of transformations of the 2nd kind in the DFM), while they do not act at all on the bare fields, which are differential forms and therefore coordinate-invariant. 
This implies that the Lorentz-invariant dressed fields transform under general coordinate transformations in the expected way.

\section{Dressing Field Method in supersymmetric field theory}\label{Dressing Field Method in supersymmetric field theory}

The DFM has also been applied to the foundations of supersymmetric field theory, in particular to the case of the Rarita-Schwinger (RS) and gravitino fields \cite{JTF-Ravera2024-SUSY,JTF-Ravera2025DFMSusyReview&Misu}: There it was shown that, 
while usually understood to result from a gauge fixing, these are actually instances of ``self-dressed'', relational variables.
This fact emerges right from the start, that is from the kinematics one typically considers in supersymmetric field theory, and has deep consequences on the formulation of theories based on supersymmetry (susy) (see, e.g., \cite{JTF-Ravera2025offshellsusy}). Moreover, the DFM is at the very root of ``unconventional supersymmetry'' (ususy) formulated via the so-called \emph{matter ansatz}, as shown in \cite{JTF-Ravera2024ususyDFM}. 
The understanding of this fact is not only foundational to the three-dimensional ususy model originally proposed in \cite{Alvarez:2011gd} (AVZ model) and to its entire physical and geometric framework, but it is also the foundation for the formulation of a novel supergeometric setup: a \emph{Matter-Interaction Supergeometric Unification} (MISU), in which the framework of supersymmetric field theory is used as a tool to provide a unified description of fermionic matter fields and bosonic gauge fields, as parts of a single superconnection, while remaining agnostic on the ultimate fate of supersymmetry and the existence of particle superpartners.
This framework is based on  supersymmetrizations of the Lorentz algebra, and the matching between bosonic and fermionic d.o.f. is not required \cite{Sohnius:1985qm}. 
Furthermore, it allows to extend the ususy idea to higher dimensions, providing an unambiguous setup with a solid foundation in differential geometry.

In this section of these notes, we review the application of the DFM to the case of RS (for simplicity, we focus on the case $\mathcal{N}=1$, $D=4$, although the approach can be extended to any number of supercharges and dimensions -- see, e.g., \cite{JTF-Ravera2025offshellsusy}). 
Here, we extract a dressing by solving a functional constraint on the RS field itself (commonly implemented as a gauge-fixing condition; this is similar to what we have already seen in the case of Maxwell theory with gauge symmetry $\U(1)$), and then revisit the intrinsically relational construction of the MISU. 
We also review, in this section, the BRST algebra and its dressed formulation, applying it to both of the cases mentioned above. 
All of this is based on the works \cite{JTF-Ravera2024-SUSY,JTF-Ravera2024ususyDFM,JTF-Ravera2025DFMSusyReview&Misu}, to which we refer the interested reader for further details (and applications of the formalism to the case of supergravity as well).

\subsection{Rarita-Schwinger spinor-vector self-dressing}\label{Rarita-Schwinger spinor-vector self-dressing}

The RS field in supersymmetric field theories is considered to be a spinor-vector ${\psi^\alpha}_{\!\mu}$, component of a spinor-valued 1-form field $\psi = \psi_\mu \, dx^{\,\mu} \in \Omega^1(U,\sf S)$, with $U \subset M$ a 4-dimensional manifold and $\sf S$ a (Dirac) spinor representation for the Lorentz group $S\!O(1,3)$. For notational convenience, we will frequently omit the spinor index $\alpha$.
We assume the Majorana condition $\bar \psi = \psi^\dagger \gamma^0 = \psi^t C$, with $C$ the charge conjugation matrix (s.t. $C^t=-C$).
In the context of supersymmetric field theory, the Lagrangian referred to as the RS term is the (massless) theory
\begin{align}
\label{RSLagr}
    L_{\text{RS}}(\psi) = \bar \psi \wedge \gamma_5 \gamma \wedge d \psi \quad \rarrow \quad
    \L_{\text{RS}}(\psi) = \epsilon^{\,\mu \nu \rho \sigma} \bar \psi_\mu \gamma_5 \gamma_\nu \d_\rho \psi_\sigma ,
\end{align}
with $\gamma:=\gamma_\mu d x^{\,\mu}$ the gamma-matrix 1-form. 
The field equations are
\begin{align}
\label{RSfieldeqs}
    \gamma_5 \gamma \wedge d \psi = 0 \quad \rarrow \quad
    \epsilon^{\,\mu \nu \rho \sigma} \gamma_5 \gamma_\nu \d_\rho \psi_\sigma = 0 .
\end{align}
The Lagrangian \eqref{RSLagr} is quasi-invariant under the susy gauge transformation
\begin{equation}
\begin{aligned}
\label{susygaugetrRS}
    & \psi \mapsto \psi^\upepsilon=\psi + d \upepsilon \quad \rarrow \quad \psi_\mu \mapsto \psi^\upepsilon_\mu=\psi_\mu + \d_\mu \upepsilon , \\
    & \text{infinitesimally, } \quad \delta_\epsilon \psi = d \epsilon \quad \rarrow \quad \delta_\epsilon \psi_\mu = \partial_\mu \epsilon ,
\end{aligned}
\end{equation}
where $\epsilon$ is the linearisation of
the Majorana spinor $\upepsilon=\upepsilon(x)$ belonging to the Abelian (additive) gauge group 
\begin{align}
\label{stranslation-gauge-grp}
\E\defeq\Big\{\upepsilon, \upepsilon'\!:\!U \rarrow T^{0|4}\, |\, \upepsilon^{\upepsilon'}=\upepsilon\, \Big\}, 
\end{align}
where $T^{0|4}\subset T^{4|4}$ is the \emph{supertranslation} subgroup  of the ($\mathcal{N}=1$) super-Poincaré group $sIS\!O(1,3) \defeq S\!O(1,3) \ltimes T^{4|4}$ \cite{Gursey1987,DeAzc-Izq}.
Notice that
\begin{align}
\left(\psi^\upepsilon\right)^{\upepsilon'}= \left( \psi + d \upepsilon \right)^{\upepsilon'} = \psi^{\upepsilon'} + d\upepsilon^{\upepsilon'} = \psi + d\upepsilon' + d \upepsilon = \psi + d (\upepsilon + \upepsilon') = \psi^{\upepsilon + \upepsilon'}.
\end{align}
We have then the susy transformation of the the Lagrangian
\begin{align}
L_\text{RS}(\psi^\upepsilon) 
= L_\text{RS}(\psi) + db(\psi; \upepsilon)
=L_\text{RS}(\psi) + d(\bar \upepsilon \wedge \gamma_5 \gamma \wedge d \psi ).
\end{align}
In the (``standard'') susy literature, it is typically considered the following ``gauge fixing'' on $\psi_\mu$:
\begin{align}
\label{gammatr}
    \gamma^{\,\mu} \psi_{\mu} = 0 .
\end{align}
We will now see that 
the condition \eqref{gammatr} can be invariantly implemented through symmetry reduction via the DFM.  
Therefore, what is commonly referred to as the (``gauge-fixed'') RS field $\psi$ is in fact obtained via the DFM as a \emph{self-dressed}, \emph{susy-invariant} variable.
Before showing this, let us just recall that taking together the field equations \eqref{RSfieldeqs} and the condition \eqref{gammatr}, one also obtains the transversality condition
\begin{align}
\label{transvcond}
    \d^{\,\mu} \psi_\mu = 0 ,
\end{align}
which therefore follows only \emph{on-shell} in the theory at hand. 
Indeed, in the \emph{flat} case, using $[\d_\mu,\gamma_\mu]=0$ together with the properties of the gamma-matrices, it can be shown that the field equations \eqref{RSfieldeqs} imply $\slashed\d (\gamma^{\,\mu} \psi_\mu)-\d^{\,\mu}\psi_\mu=0$, which, taking into account \eqref{gammatr}, yields \eqref{transvcond}. 
Using the same field equations \eqref{RSfieldeqs} together with the alternative choice \eqref{transvcond}, one gets, instead, the weaker constraint $\slashed\d (\gamma^{\,\mu}\psi_\mu)=0$.

\subsubsection{Gamma-trace dressing}\label{Gamma-trace dressing}

We will now show that the ``gamma-trace gauge fixing'' can actually be seen as an instance of the DFM. 
Let us consider the following ``gamma-trace (reducible) decomposition'':
\begin{align}\label{gammatracedec}
    {\psi^\alpha}_{\!\mu} (\uprho,\chi) := {\uprho^\alpha}_{\!\mu} + \gamma_\mu \, \chi^\alpha ,
\end{align}
where $\chi^\alpha := \tfrac{1}{4} \gamma^{\,\mu} {\psi^\alpha}_{\!\mu}$ is a spin-$1/2$ field and ${\uprho^\alpha}_{\!\mu}$ is s.t. $\gamma^{\,\mu}\uprho_\mu =0$.
We proceed point by point in our analysis.
\begin{enumerate}
    \item The susy transformation  \eqref{susygaugetrRS} of the reducible gamma-trace decomposition \eqref{gammatracedec}, in $D=4$ dimensions, is
    \begin{equation}
    \begin{aligned}
    \label{chirhogaugetr}
    \chi & \,\mapsto\, \chi^\upepsilon= \chi + \tfrac{1}{4} \slashed\d \upepsilon , \\
    \uprho_\mu & \,\mapsto \,\uprho^\upepsilon_\mu = \uprho_\mu - \tfrac{1}{4} \gamma_\mu \slashed \d \upepsilon + \d_\mu \upepsilon .
    \end{aligned}
    \end{equation}
    \item We pick the gamma-tracelessness constraint \eqref{gammatr} as a functional condition on the variable
    \begin{align}
        \psi_\mu^u\defeq \psi_\mu+ \d_\mu u
    \end{align}
    and solve it explicitly for $u$:\footnote{Let us omit, for simplicity, the spinorial object $g^\alpha$ (spin-$1/2$ field) in the kernel of $\slashed{\partial}$, s.t. $\slashed{\partial}g^\alpha=0$. In the DFM, this is a case of transformations of the 2nd kind (``ambiguity'' in the dressing field), and it can be treated analogously to what we have done in the Maxwell $\U(1)$ theory -- cf. Section \ref{Maxwell electromagnetism and the ``Lorenz gauge as a dressing'' via the DFM}. Remark that those \emph{are not} a residual of the gauge group reduced via dressing (at best, they can be an isomorphic group)). As a clarifying example, we may also consider the case of the ``tetrad dressing'', cf. Section \ref{Lorentz dressing in GR}, where transformations of the 2nd kind are simply coordinate transformations.}
    \begin{equation}
    \label{gamma-tr-dressing}
    \gamma^{\,\mu} \psi^u_\mu = \gamma^{\,\mu} (\psi_\mu + \d_\mu u) = 0 \quad \Rightarrow \quad u[\psi] = - \slashed{\d}\- (\gamma^{\,\mu} \psi_\mu) = - 4 \slashed{\d}\- \chi.
    \end{equation}
    \item We now have to assess if $u$ is 
    \begin{itemize}
        \item[i)] an element of the gauge group, and thus the gamma-trace constraint \eqref{gammatr} is a gauge fixing;
        \item[ii)] a dressing field, in which case $u$ has to transform accordingly.
    \end{itemize}
    For \eqref{gammatr} to be a gauge fixing, $u$ must be an element of the gauge group $\E$: i.e., it must be gauge-invariant, $u[\psi]^\upepsilon:=u[\psi^\upepsilon]=u[\psi]$ -- because the gauge group $\E$ of supertranslation is Abelian \eqref{stranslation-gauge-grp}. 
    But we can easily check that, as a functional of $\psi$, under the susy transformations \eqref{chirhogaugetr} $u$ gauge-transforms as
    \begin{align}
    \label{udressfieldtrepsRS1}
    u[\psi]^\upepsilon \defeq u[\psi^\upepsilon] 
    =  - 4 \slashed{\d}\- \chi^\upepsilon 
    =- 4 \slashed{\d}\- \left(\, \chi + \tfrac{1}{4} \slashed\d \upepsilon \right) 
    =- 4 \slashed{\d}\- \chi - \upepsilon
    =u[\psi] - \upepsilon.
    \end{align}
    The latter is, in fact, the Abelian (additive) version of a \emph{dressing field} transformation. 
    We thus conclude that the explicit solution of the gamma-trace constraint \eqref{gammatr} does not result in a gauge fixing but in a dressing operation, in the precise technical sense stemming from the DFM. 
    Observe that this dressing is \emph{non-local}. 
    \item The variable $\psi^u$ is thus a  susy-invariant dressed field, expressed in terms of bare fields as
    \begin{align}
    \label{dressed-RS}
    \psi_\mu^u\defeq  \psi_\mu+ \d_\mu u [\psi]
    = \psi_\mu   - 4 \d_\mu  \slashed{\d}\- \chi .
    \end{align}
    It satisfies $\gamma^{\,\mu}\psi^u_\mu \equiv 0$ by construction. 
    \item We may finally have a look at the decomposition \eqref{gammatracedec} applied to $\psi^u_\mu$. 
    We find 
    \begin{equation}
    \begin{aligned}
    \label{chirhogammatrdressed}
    \chi^u & = \chi + \tfrac{1}{4} \slashed\d u[\psi] = \chi + \tfrac{1}{4} (-4) \slashed{\d} \slashed{\d}\- \chi \equiv 0 , \\
    \uprho^u_\mu & = \uprho_\mu - \tfrac{1}{4} \gamma_\mu \slashed \d u[\psi] + \d_\mu u[\psi] = \psi_\mu^u.
    \end{aligned}
    \end{equation}
\end{enumerate}
In conclusion, in this case the dressing field is given in terms of the gamma-trace component $\chi$.
It should be stressed that in the DFM treatment, the d.o.f. of the dressed RS field \eqref{dressed-RS} are obtained in a \emph{susy-invariant} way, without any restriction on the gauge group.\footnote{The above template can be repeated by considering the alternative functional constraint \eqref{transvcond}, which is (again) usually understood as a gauge fixing. 
When solved explicitly, it again yields a dressing field as technically defined in the DFM, see \cite{JTF-Ravera2024-SUSY,JTF-Ravera2025DFMSusyReview&Misu} for details.}
The dressed field $\psi^u$ is a \emph{relational variable} \cite{JTF-Ravera2024c}: it encodes the physical, invariant relations among the d.o.f. of $\psi$. 
Observe that $\E$-invariance is achieved at the ``cost'' of locality, hinting at the fact that susy is, in this context, what is called a \emph{substantial} (or \emph{substantive}) symmetry \cite{Francois2018}.

Implementing the DFM at the level of the dynamics, 
the Lagrangian $4$-form of the dressed theory is
\begin{align}
\label{dressed-RSLagr}
    L_{\text{RS}}(\psi^u) = \bar \psi^u \wedge \gamma_5 \gamma \wedge d \psi^u 
    =L_{\text{RS}}(\psi) + db(\psi; u) =\bar \psi \wedge \gamma_5 \gamma \wedge d \psi + d(\bar u \wedge \gamma_5 \gamma \wedge d \psi ).
\end{align}
In components, we have the dressed Lagrangian density
\begin{align}
    \L_{\text{RS}}(\psi^u) = \epsilon^{\,\mu \nu \rho \sigma} \bar \psi_\mu^u \gamma_5 \gamma_\nu \d_\rho \psi^u_\sigma.
\end{align}
The dressed field equations read
\begin{align}
\label{dressed-RSfieldeqs}
    \gamma_5 \gamma \wedge d \psi^u = 0 \quad \rarrow \quad
    \epsilon^{\,\mu \nu \rho \sigma} \gamma_5 \gamma_\nu \d_\rho \psi_\sigma^u = 0.
\end{align}
The dressed Lagrangian \eqref{dressed-RSLagr} is $\E$-invariant because $\psi^u$ is a \emph{susy singlet} ($\E$-invariant). 
Given the relational, gauge-invariant character of the dressed theory, the dressed field equations \eqref{dressed-RSfieldeqs} are deterministic, meaning that, once initial conditions are specified,
they determine in a unique way the evolution of the relational d.o.f. encoded by $\psi^u$.

\subsection{BRST and dressed BRST formalisms in GFT}\label{BRST and dressed BRST formalisms in GFT}

We now review the BRST formalism in GFT, starting from its ``bare'' version and then presenting its dressed counterpart, in order to consider an example of application, specifically, to the RS model discussed just above.

\subsubsection{BRST algebra}\label{BRST algebra}

In GFT, to the \emph{infinitesimal} generators of gauge transformations, one can associate a Faddeev-Popov
\emph{ghost field} $c$, which is the field-theoretic place holder for the Maurer-Cartan form of the gauge group $\mathcal{H}$. 
The gauge transformations of the fields, $\upphi=\lbrace{A,\vphi \rbrace}$, are defined by the action of $\H$ at the finite level and of Lie$\H$ \emph{infinitesimally}. The latter reads
\begin{align}
\label{GTgauge-fields}
\delta_\lambda A = D\lambda=d\lambda +\ad(A) \lambda, \quad \delta_\lambda \vphi=-\rho_*(\lambda) \vphi, \quad \lambda \in \text{Lie}\H.
\end{align}
The \emph{BRST algebra} of a (non-Abelian) GFT with field content $\upphi=\lbrace{A,\vphi \rbrace}$ is defined as (see \cite{Becchi:1975nq})
\begin{equation}
\label{BRSTalgebradef}
\begin{aligned}
    & sA=-Dc \defeq -dc -[A,c], \quad s \vphi = -\rho_*(c)\vphi , \\
    & sc = - \tfrac{1}{2}[c,c] , 
\end{aligned}
\end{equation}
with ghost field $c$,
and we also have, for the curvature $F$ and covariant derivative of the matter field $\vphi$,
\begin{align}
\label{curvBRSTalg}
    sF = [F,c] , \quad s(D\vphi) = - \rho_*(c) D\vphi.
\end{align}
The BRST operator $s$ is an antiderivation which anticommutes with
the exterior differential $d$, so that  $sd+ds=0$, and with odd differential forms. 
The bracket $[ \,, \,]$ is  graded  w.r.t. the form and ghost degrees. It is easily verified that $s^2=0$ -- and, since $d^2=0$, we have $(s+d)^2=0$.
Let us mention, for the sake of completeness, that such differential algebra can be recast as a bigraded algebra, with total degree the sum of the form and ghost degrees, whose nilpotent operator is $\tilde{d}\defeq d+s$, s.t. $\tilde{d}^2=0$.
One defines the ``algebraic connection'' \cite{Dubois-Violette:1986vtp} $\tilde{A}\defeq A+c$, of bidegree $1$. 
The above BRST algebra follows from what Stora referred to as the ``Russian formula'' $\tilde{d}\tilde{A}+\tfrac{1}{2}[\tilde{A},\tilde{A}]=F$, by expanding it w.r.t. the ghost degree. 
The same can be done for the matter sector. 
The BRST transformation of the Lagrangian is
\begin{align}
  s L(A, \vphi) = d\beta(A, \vphi; c),
\end{align}
which reproduces the infinitesimal gauge transformation of the Lagrangian, $\delta_\lambda L(A, \vphi) = d\beta(A, \vphi; \lambda)$.
A Lagrangian functional thus belong to the $s$ modulo $d$ cohomology: $L \in H^{0,n}(s | d)$, where $0$ is the ghost degree and $n$ the de Rham form degree.

\subsubsection{Dressed BRST algebra}\label{Dressed BRST algebra}

Here we review the \emph{dressed BRST formalism} in GFT, applications of which to conformal Cartan geometry and twistor theory may be found in \cite{FLM2015_II,FLM2016_I}.

Given the definition of the ``bare'' BRST algebra \eqref{BRSTalgebradef}-\eqref{curvBRSTalg} and of the dressed fields \eqref{dressed-fields}, it is easy to show that the latter satisfy a dressed BRST algebra
\begin{equation}
\label{dressed-BRSTalgebradef}
\begin{aligned}
    & sA^u=-D^u c^u , \quad s \vphi^u = -\rho_*(c^u)\vphi^u, \quad 
     sF^u = [F^u,c^u] , \quad s(D^u\vphi^u) = - \rho_*(c^u) D^u\vphi^u, \\
    & sc^u = - \tfrac{1}{2}[c^u,c^u],
\end{aligned}
\end{equation}
with the \emph{dressed ghost} $c^u$ defined as
\begin{align}
    \label{dressed-ghost}
    c^u \defeq u\- c u + u\- s u .
\end{align}
This is formal, leaving the BRST variation $su$ of the field $u$ unspecified, i.e., independent of whether or not $u$ is a dressing field. 

If $u$ is a $\H$-dressing field indeed, then it satisfies the BRST version of the defining property of a dressing field, namely
\begin{align}
    su=-cu.
\end{align}
So, the dressed ghost is 
\begin{align}
    c^u \defeq u\- c u + u\- s u = u\- c u + u\- (-cu) \equiv 0,
\end{align}
which trivializes the dressed BRST algebra:
\begin{align}
    sA^u = 0 , \quad s \vphi^u =0 , \quad sF^u=0 , \quad s(D^u \vphi^u) =0,
\end{align}
thereby expressing that the $\H$ symmetry is totally reduced.
In other words, invariant, dressed variables $\upphi^u$ are in the kernel of the differential BRST operator $s$. 
Therefore, the dressed Lagrangian is $s$-closed: $sL(A^u,\vphi^u)=0$.

If, on the other hand, $u$ is a $\K$-dressing field -- as we have previously discussed -- one expects the residual $\J$-gauge symmetry of the 1st kind. We may then write
\begin{equation}
\label{splittingresghostands}
\begin{aligned}
    & c = c_K + c_J , \\
    & s = s_K + s_J.
\end{aligned}
\end{equation}
From the defining property $s_K u=-c_K u$ of the dressing field, it follows that the dressed ghost is
\begin{equation}
\label{dressed-ghost-res1stkind}
\begin{aligned}
    c^u & = u\- (c_K+c_J) u + u\- s u \\
    & = u\- c_K u + u\- c_J u + u\-(-c_K u) + u\- s_J u \\
    & = u\- c_J u + u\- s_J u,
\end{aligned}
\end{equation}
so that the dressed BRST algebra encodes the residual $\J$-symmetry of the $\K$-invariant dressed fields $\upphi^u$.
The explicit form of $s_J u$ depends on the specific case under analysis.
If, for example, the $\K$-dressing field transforms as $u^\eta =\eta\- u\, \eta$ for $\eta\in \J$, so that the corresponding BRST variation is $s_J u = [u,c_J]$, we see that the dressed ghost \eqref{dressed-ghost-res1stkind} simply reduces to $c^u = c_J$, so that  \eqref{dressed-BRSTalgebradef}  becomes
\begin{equation}
\begin{aligned}
    & sA^u=-D^u c_J , \quad s \vphi^u = -\rho_*(c_J)\vphi^u , \\
    & sF^u = [F^u,c_J] , \quad s(D^u\vphi^u) = - \rho_*(c_J)D^u\vphi^u, \\
    & sc_J = - \tfrac{1}{2}[c_J,c_J] .
\end{aligned}
\end{equation}
This shows that the $\K$-invariant dressed fields are standard $\J$-gauge fields: 
on them the BRST operator reduces to $s=s_J$. 
Then, the BRST variation of the dressed Lagrangian is $sL(A^u,\vphi^u)=s_J L(A^u,\vphi^u)=d\beta(A^u,\vphi^u;c_J)$, i.e., it belongs to the $s_J$ modulo $d$ cohomology: $L^u \in H^{0|n}(s_J|d)$.\footnote{We refer the interested reader to the appendix of \cite{JTF-Ravera2025DFMSusyReview&Misu} for the perturbative version of the above, i.e., the \emph{perturbatively dressed BRST algebra} satisfied by perturbatively dressed fields.}

\subsubsection{Dressed BRST formalism for the RS case}\label{Dressed BRST formalism for the RS case}

Let us now introduce the BRST operator $s_{\text{susy}}$ associated with supersymmetry, and the BRST algebra for the RS field,
\begin{equation}
\begin{aligned}
    & s_{\text{susy}} \psi = - dc , \\
    & s_{\text{susy}} c = 0 ,
\end{aligned}
\end{equation}
where $c$ is a spinorial ghost field and its BRST transformation identically vanishes due to the additive Abelian character of the symmetry group.
This BRST algebra simply encodes the (infinitesimal) susy transformations \eqref{susygaugetrRS}. 
For the gamma-trace dressing\footnote{And, alternatively, for the ``transverse dressing'', obtained by solving explicitly $\partial^{\,\mu}\psi^u_\mu=0$ for $u$, as well \cite{JTF-Ravera2024-SUSY,JTF-Ravera2025DFMSusyReview&Misu}.} we have 
\begin{align}
    s_{\text{susy}} u = -c , 
\end{align}
which is
the BRST version of \eqref{udressfieldtrepsRS1}. Correspondingly, the dressed (Abelian) ghost is  
\begin{align}
    c^u= c + s_{\text{susy}} u = c -c =0,
\end{align}
which implies the triviality of the dressed BRST algebra:
\begin{equation}
\begin{aligned}
    & s_{\text{susy}} \psi^u = - dc^u = 0 , \\
    & s_{\text{susy}} c^u = 0 ,
\end{aligned}
\end{equation}
i.e., the susy-invariance of $\psi^u$, as expected.  
One can explicitly verify the BRST invariance of $\psi^u$. 
Indeed, we have
\begin{equation}
\begin{aligned}
    & u = u[\psi] = - 4 \slashed{\d}\- \chi \\
    &\Rightarrow \quad s_{\text{susy}} \psi^u_\mu = s_{\text{susy}} ( \psi_\mu - 4 \partial_\mu \slashed{\d}\- \chi ) = s_{\text{susy}} \psi_\mu - 4 \partial_\mu \slashed{\d}\- ( s_{\text{susy}} \chi ) = - \partial_\mu c + \partial_\mu c = 0 ,
\end{aligned}
\end{equation}
where we have used the fact that $s_{\text{susy}} \chi = - \tfrac{1}{4}\slashed{\d} c$, the BRST version of the first line of \eqref{susygaugetrRS}.
The dressed Lagrangian $L_{\text{RS}}(\psi^u)$ satisfies $s_{\text{susy}} L_{\text{RS}}(\psi^u) \equiv 0$, as susy is completely reduced, and the dressed Lagrangian is written in terms of susy singlets.
Remark that, in the presence of another gauge group $\J$ besides susy, one should consider $s=s_{\text{susy}} + s_J$, where $s_J$ is the BRST operator associated with the residual $\J$-transformations.
In this case, after susy reduction, the dressed ghost depends on $c_J$ and $s_J u$ -- see eq. \eqref{dressed-ghost-res1stkind} -- so that the dressed BRST algebra \eqref{dressed-BRSTalgebradef} encodes the linear $\J$-transformation of $\psi^u$.

\subsection{Relational Matter-Interaction Supergeometric Unification via the DFM}\label{Matter-Interaction Supergeometric Unification via the DFM}

In this section, we review
a novel Matter-Interaction Supergeometric Unification (MISU) scheme, inspired by the analysis of ``unconventional supersymmetry'' (ususy) via the DFM done in \cite{JTF-Ravera2024ususyDFM}. 
We shall illustrate this general template by treating the simplest MISU kinematics based on the Lorentz superalgebra, and deriving its dressed BRST algebra.\footnote{This topic was not covered in person during the conference school due to time constraints; I therefore provide the details here, reviewing the derivation of the formalism and its application to an illustrative example, for the interested reader.} 

Supersymmetric field theory as conventionally applied within high-energy physics implies a doubling of the number of fundamental fields (and particles), whereby each known particle has a \emph{superpartner} of opposite statistics: fermionic (matter) fields have bosonic partners, bosonic (gauge and Higgs) fields have fermionic partners. 
This view can only be accommodated with empirical data via
the notion of (spontaneous) susy breaking.
Still, until now, supersymmetric particles did not show up in colliders. 
However, this fact does not decisively undermine the framework of differential supergeometry (the mathematical foundation of supersymmetric field theory), which does not require a matching of bosonic and fermionic d.o.f. \cite{Sohnius:1985qm}.
The MISU framework exploits supergeometry without the d.o.f. matching constraint to describe gauge interactions and matter fields as parts of the same superconnection. 
It is thus closer to Berezin's original motivation for the introduction of supergeometry in fundamental
physics \cite{Berezin-Marinov1977}.
As we will see, the DFM is an integral part of MISU, so that the unifying invariant superconnection is fundamentally a relational variable.
Here, let us sketch the template for a MISU kinematics.

In MISU, 
the kinematics is given by
a superbundle 
$P\rarrow M$, where the base $M$ is bosonic, and whose structure group $H$ is a graded Lie group: correspondingly, the gauge supergroup is $\H$, i.e., susy is ``\emph{internal}''. 
The full supergroup of local transformations is $\Diff(M)\ltimes \H$.
The field space on which it acts is $\Phi=\{e, \mathbb{A}\}$, where $e=e^a={e^a}_\mu dx^{\,\mu}$
is the canonical soldering form of $M$ and $\mathbb{A}$ is a (non-canonical) $\LieH$-valued Ehresmann superconnection -- $\LieH$ is the Lie superalgebra of $H$.
There are two options for choosing $\LieH$ in a MISU kinematics:
\begin{itemize}
    \item One may simply select the Lie superalgebra  $\LieH$ s.t. the bracket of its susy generators $\mathbb Q$ is not generated by  ``(internal) infinitesimal translation generators'' $\mathbb P$.
    In this way, there is no spurious ``internal translation potential'' in $\mathbb{A}$, which would be redundant (or would need to be identified) with the already existing soldering $e$ of $M$.

    \item Less trivially, one may consider Cartan supergeometries, see, e.g., \cite{JTF-Ravera2024review,JTF-Ravera2024ususyDFM}, modeled on Klein pairs of Lie superalgebras $(\mathfrak{g}, \mathfrak{h})$, s.t. $\mathfrak{h}\subset \mathfrak{g}$, and $\mathfrak{g}/\mathfrak{h}$ is a bosonic {$H$-module} with basis (generators) $\mathbb P$ (infinitesimal ``translations'').
    Then, $P\rarrow M$ is a superbundle, as required above, and we have a Cartan geometry $(P, \b{\mathbb{A}})$ with $\LieG$-valued Cartan superconnection $\b{\mathbb{A}}$ splitting as a $\LieH$-valued Ehresmann superconnection $\mathbb{A}$ and a  $(\mathfrak{g}/\mathfrak{h})$-valued soldering form (vielbein) $e=e^a$; i.e., the field space is the space of Cartan connections $\Phi=\{\b{\mathbb{A}} \}$, which is naturally subject to internal transformations under $\Diff(M)\ltimes \H$ (and the latter contains no ``gauged translations'').  
\end{itemize}
To find  candidates suiting either of the above desiderata, for $\LieH$ and/or $(\LieG, \LieH)$, one pick from the classification of superalgebras in, e.g., \cite{Kac1977}.

In either of the two options, to obtain the  MISU kinematics one must apply the DFM:
In both cases, the susy part of the superconnection $\mathbb A$ is indeed a spinor 1-form field $\psi=\psi^\alpha$, from which a susy-dressing field $u[\mathbb A]$ can be extracted via a gamma-trace decomposition.
One thus obtains a susy-invariant dressed superconnection $\mathbb{A}^u$, whose susy component is of the form $\gamma \, \chi^u = dx^{\,\mu}\gamma_\mu \,\chi^u$, with $\chi^u$ a spinor field, potentially describing a \emph{matter field} -- and $\gamma=dx^{\,\mu}\gamma_\mu = dx^{\,\mu}\gamma_a {e^a}_\mu$ the gamma matrix 1-form --
and whose bosonic component 
describes a gauge potential.
Thus, the  susy-invariant relational superconnection $\mathbb A^u$ geometrically (kinematically) unifies gauge and matter fields. 

Given a MISU kinematics obtained following the above template, many models may be built, i.e., a dynamics may be chosen by writing down a Lagrangian. 
Model building can be approached in two ways:
\begin{itemize}
    \item One option is to propose a Lagrangian that is (quasi-)invariant under either the initial full  supergroup $\H$ -- that is, covariant under $\Diff(M) \ltimes \H$ -- i.e., to apply the Gauge Principle to the bare (``pre-MISU'') kinematics, and then dress it. 
    One thereby gets a dressed Lagrangian, from which dressed field equations for the dressed fields can be derived.\footnote{This is the case of the so-called AVZ model of ususy.}
    \item Alternatively, one may start directly from the MISU kinematics, whose residual gauge transformations of the 1st kind is the bosonic gauge group $\H/$susy, i.e., whose group of local transformations is $\Diff(M) \ltimes (\H/$susy$)$. 
    Thus, one may apply the Gauge Principle to the residual gauge transformations of the 1st kind, proposing a Lagrangian (quasi-)invariant under $\H/$susy, whose field equations for the dressed field are therefore ($\H/$susy)-covariant.
\end{itemize}
The first approach is more constraining than the second, as the gauge group of the latter is only a subgroup of the gauge group of the former. 

\subsubsection{An illustrative model: MISU based on the Lorentz superalgebra}\label{An illustrative model: MISU based on the Lorentz superalgebra}

Let us now illustrate the above template with a simple example of MISU kinematics in $D=4$.
We consider the semi-direct extension $\LieH=\mathfrak{sl}(2, \CC) \oplus \CC^2$ of the Lorentz algebra $\mathfrak{spin}(1,3)\simeq\mathfrak{sl}(2, \CC)$ by an (additive) Abelian superalgebra. 
We take a superbundle $P\rarrow M$ with structure supergroup $H=S\!L(2, \CC) \ltimes \CC^2$ over an even manifold $M$, with corresponding gauge supergroup $\H:= \{ g, g':  M \rarrow H\,|\, {g}^{\prime\, g} =g\- g' g \}$ and  gauge algebra
Lie$\H:= \{ \lambda, \lambda': M \rarrow \LieH\,|\, \delta_\lambda \lambda' = [\lambda', \lambda] \}$. 
Relying on a compact matrix notation, we may write an element of $\H$ as
\begin{align}
    g=\begin{pmatrix} B & \upepsilon\\ 0 & 1 \end{pmatrix},
\end{align}
with $B\in \SL(2, \CC)$ and $\upepsilon:M \rarrow  \CC^2$ a susy spinor; the semi-direct structure of $\H$ is reproduced by matrix multiplication.
Let us also introduce the notation 
\begin{align}
  \mathbb{B} \defeq \begin{pmatrix} B & 0\\ 0 & 1 \end{pmatrix} , \quad  \boldsymbol{\upepsilon} \defeq  \begin{pmatrix} \mathds{1} & \upepsilon \\ 0 & 1 \end{pmatrix} 
\end{align}
for the $\SL(2, \CC)$ and the susy gauge subgroups elements, respectively.
Correspondingly, an element of Lie$\H$ is
\begin{align}
\label{gaugeparam}
\lambda 
&= \begin{pmatrix}
   \beta  & 
 \varepsilon \\
0  &\    0
    \end{pmatrix} ,
\end{align}
with $\beta \in \mathfrak{sl}(2, \CC)$ and $\varepsilon \in \CC^2$ a spinorial susy parameter. 
The Ehresmann superconnection and its curvature are thus
\begin{equation}
\begin{aligned}
\label{connection-ex1}
    \mathbb{A} 
= \begin{pmatrix}
  \omega &\   \psi \\
   0  & \  0
\end{pmatrix}
\quad \text{and} \quad\ 
\mathbb{F} =d \mathbb{A}  + \tfrac{1}{2}[\mathbb{A} , \mathbb{A} ] = d \mathbb{A} + \mathbb{A}^2
= \begin{pmatrix}
  \Omega &\   \nabla \psi \\
   0  & \  0
\end{pmatrix}
=\begin{pmatrix}
  d\omega+\omega^2 &\   d\psi+ \omega \psi \\
   0  & \  0
\end{pmatrix},
\end{aligned}
\end{equation}
where $\nabla$ denotes the Lorentz-covariant derivative.\footnote{Again, we omit the explicit (rigid, Lorentz) indices notation on differential forms, to keep the presentation as clean and clear as possible.}
The curvature satisfies Bianchi identity $D^\mathbb{A}\mathbb{F}= d\mathbb{A} +[\mathbb{A}, \mathbb{F}]=0$.
The $\H$-transformations of $\mathbb{A}$ and $\mathbb{F}$ are
\begin{equation}
\label{GT-A-F-MISU}
\begin{aligned}
\mathbb A^g&=g\- \mathbb A g +g\-dg 
 = \begin{pmatrix}
       B\-\omega B + B\- dB & B\- (\psi+  \nabla \upepsilon ) \\ 0 & 0 
  \end{pmatrix} , \\[2mm]
   \mathbb F^{\,g}&=g\- \mathbb F g 
   = \begin{pmatrix}
       B\-\Omega B  & B\- (\nabla \psi +   \Omega\upepsilon ) \\ 0 & 0 
   \end{pmatrix},
\end{aligned}
\end{equation}
that is, we have for our elementary fields $\omega$ and $\psi$,
\begin{equation}
\begin{aligned}
    \omega^B &= B^{-1} \omega B + B^{-1} dB \quad &&\text{and} \quad \  
    \omega^\upepsilon = \omega , \\
    \psi^B &= B^{-1}  \psi  \quad &&\text{and} \quad\ 
    \psi^\upepsilon = \psi + \nabla \upepsilon .
\end{aligned}
\end{equation}
Notice that here $\omega$ is a susy singlet already.
The corresponding Lie$\H$-transformations  are
\begin{equation}
\label{linear-GT-A-F}
\begin{aligned}
    \delta_\lambda \mathbb A = D^\mathbb{A}\lambda
    =\begin{pmatrix}
      \nabla \beta  & \ -\beta \psi+ \nabla \varepsilon  \\ 0 & 0 
  \end{pmatrix}
  \quad \text{ and } \quad 
  \delta_\lambda \mathbb F = [\mathbb{F}, \lambda]
     =\begin{pmatrix}
      \left[\Omega, \beta\right]  & \ -\beta \nabla \psi + \Omega \varepsilon  \\ 0 & 0 
  \end{pmatrix},
\end{aligned}
\end{equation}
with $\nabla \beta = d\beta + [\omega ,\beta]$ and $\nabla \varepsilon = d\varepsilon + \omega \varepsilon$.
Again, notice that $\delta_\varepsilon \omega = 0$.

One may then introduce the  ``gamma-trace'' decomposition for $\psi=\psi_\mu dx^{\,\mu}$ precisely as in \eqref{gammatracedec}:  
$\psi =\uprho + \gamma \chi$, 
where, by definition, $\gamma^{\,\mu} \uprho_\mu=0$ and $\chi:= \frac{1}{4} \gamma^{\,\mu} \psi_\mu$. 
From it, we will extract a susy dressing field $u=u[\psi]$.
Under (finite) susy transformations, we have 
\begin{align}
    \psi^\upepsilon_\mu = \uprho^\upepsilon_\mu + \gamma_\mu \,\chi^\upepsilon
    \quad 
    \Rightarrow 
    \quad
    \left\{ 
    \begin{matrix}
     \ \ \uprho^\upepsilon_\mu = \uprho_\mu - \tfrac{1}{4} \gamma_\mu \slashed \nabla \upepsilon + \nabla_\mu \upepsilon ,
     \\[2mm]
     \hspace{-12mm} \chi^\upepsilon= \chi + \tfrac{1}{4} \slashed\nabla \upepsilon.
    \end{matrix}
    \right.
\end{align}
Let us now define the (field-dependent) operator 
\begin{align}
    b_\mu =b_\mu[\omega] \defeq \tfrac{1}{4} \gamma_\mu \slashed \nabla  - \nabla_\mu  ,
\end{align}
with formal left inverse $[b\-]^\mu$,
so that we may write
\begin{align}
    \uprho^\upepsilon_\mu = \uprho_\mu - b_\mu (\upepsilon).
\end{align}
Now, we consider the variable
\begin{align}
    \psi^u_\mu \defeq \psi_\mu + \nabla_\mu u= \uprho^u_\mu + \gamma_\mu \,\chi^u , 
    \quad 
    \text{satisfying the functional constraint }
        \uprho^u_\mu = 0. 
\end{align}
By solving explicitly for $u$, we extract the \emph{field-dependent dressing field}
\begin{align}
\label{drfieldmisu}
    u[\psi, \omega]=u[\mathbb A] \defeq [b^{-1}]^{\,\mu} (\uprho_\mu),
\end{align}
whose  dependence on $\omega$ comes from $b_\mu$.
Its susy transformation is in fact easily found to be
\begin{align}
\label{Abelian-misu-dressing}
    u^\upepsilon = u[\mathbb A^\upepsilon] = u[\psi^\upepsilon,\omega^\upepsilon] = u[\psi^\upepsilon,\omega] = (b[\omega]^{-1})^{\,\mu} (\uprho^\upepsilon_\mu) = (b[\omega]^{-1})^{\,\mu} (\uprho_\mu - b_\mu (\upepsilon)) = u[\mathbb A] - \upepsilon.
\end{align}
We have thus a \emph{non-local}, Abelian dressing field.
We can also use the matrix notation
\begin{align}
\boldsymbol{u}=
\begin{pmatrix}
\mathds{1} & \  u \\
0 & 1
\end{pmatrix}=\begin{pmatrix}
\mathds{1} & \  u[\mathbb A] \\
0 & 1
\end{pmatrix},
\end{align}
so that the standard defining property of a dressing fields is
\begin{align}
\label{susy-GT-u}
\boldsymbol{u}^{\boldsymbol{\upepsilon}} = {\boldsymbol{\upepsilon}}^{-1} \boldsymbol{u} =\begin{pmatrix}
\mathds{1} & \  -\upepsilon \\
0 & 1
\end{pmatrix}
\begin{pmatrix}
\mathds{1} & \  u \\
0 & 1
\end{pmatrix}= \begin{pmatrix}
\mathds{1} & \  u - \upepsilon \\
0 & 1
\end{pmatrix},
\end{align}
reproducing the Abelian transformation \eqref{Abelian-misu-dressing}. 
Thus, with the dressing field $\boldsymbol{u}$ we can build the susy-invariant superconnection
\begin{align}
\label{dressed-A}
\mathbb{A}^{\boldsymbol{u}} \defeq \boldsymbol{u}\- \mathbb{A} \boldsymbol{u} + \boldsymbol{u}\- \, d \boldsymbol{u} = \begin{pmatrix}
 \ \omega^u &\  \psi^u \\
   0  &\ 0
\end{pmatrix} = \begin{pmatrix}
 \ \omega &\  \psi + \nabla u \\
   0  &\ 0
\end{pmatrix}
=
\begin{pmatrix}
 \ \omega &\  \gamma \, \chi^u \\
   0  &\ 0
\end{pmatrix} ,
\end{align}
with  the susy-invariant dressed spinor $\chi^u \defeq \chi + \tfrac{1}{4} \slashed{\nabla} u$, by construction. 
Therefore, gauge and matter fields now feature on equal footing in the dressed superconnection $\mathbb A^{\boldsymbol{u}}$. 
The corresponding dressed curvature is
\begin{align}
\label{dressed-F}
\mathbb{F}^{\,\boldsymbol{u}} = \begin{pmatrix}
     \Omega^u &\  \gamma_a T^a \chi^u - \gamma \,\nabla \chi^u \\ 
     0 & 0
\end{pmatrix} = \begin{pmatrix}
     \Omega &\ \gamma_a T^a \chi^u - 
 \gamma \, \nabla \chi^u \\ 
     0 & 0
\end{pmatrix},
\end{align}
where $\Omega^u=\Omega$, and $T^a\defeq de^a + {\omega^a}_b e^b$ is the torsion of $M$.
Let us remark that, since $u=u[\mathbb A]$, the susy-invariant dressed field $\mathbb A^{\boldsymbol{u}}$, and, in particular, $\psi^u=\gamma \chi^u$, is a  self-dressed \emph{relational} field variable. 

Note that 
the dressed fields $\mathbb A^{\boldsymbol{u}}$ and $\mathbb F^{\,\boldsymbol{u}}$ are expected to exhibit residual $\SL(2, \CC)$-transformations (namely, we are in the presence of a residual symmetry of the 1st kind, in the DFM language). 
To find them, one only needs to determine the $\SL(2, \CC)$-transformations of the dressing, that is $\boldsymbol{u}^\mathbb B$.  
We have
\begin{equation}
\label{restr1stkinfmisu}
\begin{aligned}
u^B  \defeq&\, u[\mathbb A^B] 
= (b[\omega^B]^{-1})^{\,\mu} (\uprho^B_\mu) 
= B\-b[\omega]^{-1} B (B\- \uprho_\mu) 
= B\- b[\omega]^{-1} (\uprho_\mu) \\
=&\, B\- u[\mathbb{A}],
\end{aligned}
\end{equation}
where we used the fact that $b_
\mu[\omega]$ is a Lorentz-covariant operator. 
This can be cast in matrix form as
\begin{align}
\label{Lorentz-GT-u}
\boldsymbol{u}^{\mathbb{B}} 
 = \mathbb{B}\- \boldsymbol{u}\, \mathbb{B} 
  = 
\begin{pmatrix} B\- & 0\\ 0 & 1 \end{pmatrix}
\begin{pmatrix}
\mathds{1} & \  u[\mathbb A] \\
0 & 1
\end{pmatrix}
\begin{pmatrix} B & 0\\ 0 & 1 \end{pmatrix}
  =
\begin{pmatrix}
\mathds{1} & \  B\- u[\mathbb A] \\
0 & 1
\end{pmatrix}. 
\end{align}
Then, since  $\mathbb{A}^{\mathbb B}$ is given by \eqref{GT-A-F-MISU} (specialising $g\rarrow \mathbb{ B}$), it is immediate that   
\begin{equation}
\begin{aligned}
\label{B}
(\mathbb{A}^{\boldsymbol{u}})^{\mathbb{B}} &=  \mathbb{B}\- \mathbb{A}^{\boldsymbol{u}}\,  \mathbb{B} + \mathbb{B}\- \, d\,  \mathbb{B} =\begin{pmatrix}
     B\-\omega B + B\- \,dB &\ B\- \gamma \, \chi^u \\ 0 & 0
 \end{pmatrix}, \\[2mm]
(\mathbb{F}^{\,\boldsymbol{u}})^{\mathbb{B}}&= \mathbb{B}\- \mathbb{F}^{\,\boldsymbol{u}}\, \mathbb{B} = \begin{pmatrix}
     B\-\Omega B &\ B\- {(\gamma_a T^a \chi^u - 
 \gamma \, \nabla \chi^u )} \\ 0 & 0
 \end{pmatrix} ,
\end{aligned}
\end{equation}
that is, the susy-invariant dressed field $\mathbb A^{\boldsymbol{u}}$, and its associated dressed curvature, have standard residual Lorentz gauge transformations.

\subsubsection{Dressed BRST formulation of the MISU model}
\label{Dressed BRST formulation of the MISU model}

The MISU framework just described can also be recast in the BRST formulation, in which one can derive the dressed BRST algebra. Here let us review this for the four-dimensional example we just presented.
The BRST algebra of the simple model above, reproducing the linear gauge transformation \eqref{linear-GT-A-F}, is 
\begin{equation}
\begin{aligned}
    s \mathbb{A} = - D^\mathbb{A} \mathbb{C},
    \quad 
    s \mathbb{F} = \left[\mathbb{F} , \mathbb{C}\right],
\end{aligned}
\end{equation}
where the ghost field splits as $\mathbb{C} = \CC_{\text{L}} + \CC_{\text{susy}}$, with a  Lorentz ghost and a susy ghost. In matrix notation,
\begin{align}
    \mathbb{C} = \CC_{\text{L}} + \CC_{\text{susy}} = \begin{pmatrix}
    \beta & \  0 \\
    0 & 0
    \end{pmatrix} + \begin{pmatrix}
    0 & \  \varepsilon \\
    0 & 0
    \end{pmatrix} = \begin{pmatrix}
    \beta & \  \varepsilon \\
    0 & 0
    \end{pmatrix},
\end{align}
with  $\beta$ and $\varepsilon$ now being assigned ghost degree 1. 
The BRST operator splits accordingly as $s = s_{\text{L}} + s_{\text{susy}}$.
The BRST versions of the defining  susy transformation  \eqref{susy-GT-u} of the dressing  $\boldsymbol{u}[\mathbb A]$, and its $\SL(2, \CC)$ transformation \eqref{Lorentz-GT-u}, read, respectively, 
\begin{align}
 s_{\text{susy}} \boldsymbol{u} = -\CC_{\text{susy}} \boldsymbol{u} 
     \quad \text{and} \quad 
 s_{\text{L}} \boldsymbol{u} = [\boldsymbol{u}, \CC_{_{\text{L}}}]. 
\end{align}
Thus, the dressed BRST algebra satisfied by $\mathbb{A}^{\boldsymbol{u}}$ and $\mathbb{F}^{\,\boldsymbol{u}}$ is 
\begin{equation}
\begin{aligned}
    s \mathbb{A}^{\boldsymbol{u}} = - D^{\mathbb{A}^{\boldsymbol{u}}} \mathbb{C}^{\boldsymbol{u}},
    \quad \text{and} \quad 
    s \mathbb{F}^{\,\boldsymbol{u}} = \left[\mathbb{F}^{\,\boldsymbol{u}} , \mathbb{C}^{\boldsymbol{u}} \right],
\end{aligned}
\end{equation}
with the dressed ghost $\CC^{\boldsymbol{u}}$ found to be
\begin{equation}
\begin{aligned}
    \CC^{\boldsymbol{u}} &= \boldsymbol{u}\- \CC \,\boldsymbol{u} + \boldsymbol{u}\- s \boldsymbol{u}\\
    & = \boldsymbol{u}\- (\CC_{\text{L}}+\CC_{\text{susy}}) \boldsymbol{u} + \boldsymbol{u}\- (s_\text{L} \boldsymbol{u} + s_{\text{susy}} \boldsymbol{u}) \\
    & = \boldsymbol{u}\- (\CC_{\text{L}} + \CC_{\text{susy}}) \boldsymbol{u} + \boldsymbol{u}\- ([\boldsymbol{u},\CC_{\text{L}}] - \CC_{\text{susy}} \boldsymbol{u}) \\
    & = \boldsymbol{u}\- \CC_{\text{L}} \boldsymbol{u} + \boldsymbol{u}\- (\boldsymbol{u} \CC_{\text{L}} - \CC_{\text{L}} \boldsymbol{u}) = \boldsymbol{u}\- \boldsymbol{u} \CC_{\text{L}} 
    = \CC_{\text{L}}.
\end{aligned}
\end{equation}
This shows that the susy-invariant dressed fields $\mathbb{A}^{\boldsymbol{u}}$ and $\mathbb{F}^{\,\boldsymbol{u}}$ are standard Lorentz gauge fields. Indeed, we have
\begin{equation}
\begin{aligned}
    &s \mathbb{A}^{\boldsymbol{u}} 
    =
    s_{\text{L}}\mathbb{A}^{\boldsymbol{u}}
    =
    - D^{\mathbb{A}^{\boldsymbol{u}}} \mathbb{C}_{\text{L}} = \begin{pmatrix}
    - \nabla \beta & \  - \gamma \,\beta \, \chi^u \\
    0 & 0
    \end{pmatrix} , \\
    & \text{and} \quad 
    s \mathbb{F}^{\,\boldsymbol{u}}
     =
    s_{\text{L}}\mathbb{F}^{\,\boldsymbol{u}}
    =
     \left[\mathbb{F}^{\,\boldsymbol{u}} , \mathbb{C}_{\text{L}} \right] = \begin{pmatrix}
    \left[\Omega, \beta \right] & \  - \gamma \, \beta \,\nabla \chi^u \\
    0 & 0
    \end{pmatrix}.
\end{aligned}
\end{equation} 
This is just kinematics. Regarding the dynamics, we remark that one now only has to require a dressed Lagrangian $L^{\boldsymbol{u}}$ to be invariant under $s_{\text{L}}$ -- the dressed fields being in the kernel of $s_{\text{susy}}$, so will be a Lagrangian functional written for them.

We conclude this section with the observation that the MISU approach encompasses the so-called ``unconventional supersymmetry'' introduced by Alvarez, Valenzuela, and Zanelli in the guise of a three-dimensional model
\cite{Alvarez:2011gd,Alvarez:2013tga}. 
In the AVZ model, based on the superalgebra $\mathfrak{osp}(2|2)$, a superconnection $\mathbb A_{\text{AVZ}}$ is chosen such that its susy part is \emph{required} to satisfy what has been dubbed the ``matter ansatz'' in \cite{Alvarez:2021zhh}: $\psi\defeq \gamma\chi$. 
The field $\mathbb A_{\text{AVZ}}$ thus encodes both a gauge and a matter field, and the model has been shown to be applicable to the description of graphene systems, see, e.g., \cite{Iorio:2014nda,Iorio:2018agc,Ciappina:2019qgj,Acquaviva:2022yiq}. 
Im \cite{JTF-Ravera2024ususyDFM}, the matter ansatz was shown to be a case of the DFM: 
In a way very similar as  above, a \emph{perturbative} susy dressing field is extracted from the general $\mathfrak{osp}(2|2)$ superconnection $\mathbb{A}_\mathfrak{osp}$, again via a gamma-trace decomposition of its odd component $\psi$. 
The perturbatively dressed connection then reproduces the AVZ ansatz with the benefit of achieving susy-invariance at first order.
We refer the reader to \cite{JTF-Ravera2024ususyDFM,JTF-Ravera2025DFMSusyReview&Misu} for further details on this topic.

\section{Dressing Field Method for general-relativistic theories}\label{Dressing Field Method for general-relativistic theories}

This section is devoted to the application of the DFM to the case of general-relativistic theories, where, remarkably, we introduce the notion of ``dressed regions'', and refine the notion of ``physical spacetime'' itself.
We refer the reader to \cite{JTF-RaveraNoBdyPb2025,JTF-Ravera2024gRGFT,Berghofer:2025ius} for further details on the formalism and applications (see also \cite{Francois:2025ptj} for an application of the DFM for the case of diffeomorphisms in the context of galaxy dynamics).

\subsection{Dressed kinematics for general-relativistic theories}\label{Dressed kinematics for general-relativistic theories}

Consider a general-relativistic theory with field content $\upphi=\{A, \vphi, g\}$, where $g$ is a metric field on $M$, supporting the pullback action of the group of diffeomorphisms, 
\begin{equation}
\label{Diff-trsf-fields}
\begin{aligned}
    & \upphi^\psi := \psi^* \upphi, \quad
    \psi \in \Diff(M),
    \\
    & \text{i.e.,} \quad \lbrace{ A^\psi, \varphi^\psi, g^\psi \rbrace} := \lbrace{ \psi^* A, \psi^* \varphi,\psi^* g \rbrace}.
\end{aligned}
\end{equation}
A \emph{dressing field for diffeomorphisms} is a smooth map
\begin{align}
\label{diffeo-dressing-field}
    \upsilon: N \rightarrow M, \quad \text{s.t.} \quad \upsilon^\psi := \psi\- \circ \upsilon, 
\end{align}
$\forall \psi \in \Diff(M)${, $N$ being a model smooth manifold (s.t. dim$N=$dim$M$, typically $N=\mathbb{R}^n$).} 
As in the case of GFT, a dressing field should be extracted from the field content of the theory  
i.e., it should be a \emph{field-dependent dressing field}, $\upsilon = \upsilon[\upphi]$, so that $\upsilon^\psi := \upsilon (\psi^* \upphi) = \psi\- \circ \upsilon[\upphi]$.
Given such a dressing field $\upsilon$, \emph{dressed fields} are defined as
\begin{equation}
\label{diffeodressedfields}
\begin{aligned}
    & \upphi^\upsilon := \upsilon^* \upphi , \\
    & \text{i.e.,} \quad \lbrace{ A^\upsilon, \varphi^\upsilon, g^\upsilon \rbrace} = \lbrace{ \upsilon^* A, \upsilon^* \varphi,\upsilon^* g \rbrace},
\end{aligned}
\end{equation}
and they are $\Diff(M)$-invariant by construction. 
We use, again, the DFM rule of thumb, here for  the case of diffeomorphisms.
For $\upsilon=\upsilon[\upphi]$, the dressed fields \eqref{diffeodressedfields} are manifestly relational variables: 
they represent $\Diff(M)$-invariant \emph{relations} among the physical spatiotemporal d.o.f. embedded in $\upphi$. 
They thus implement the first insights stemming from the point-coincidence argument (see Section \ref{Hole and point-coincidence arguments in a nutshell}).
Regarding the second insight, we 
observe that the dressed fields \eqref{diffeodressedfields}  are \emph{not} objects defined on the ``bare'' manifold $M$.
They live on \emph{dressed regions}, defined below.

\subsection{Dressed regions}\label{Dressed regions}

The dressed fields \eqref{diffeodressedfields} live on 
field-dependent \emph{dressed regions} defined by
\begin{align}
\label{fielddepdrreg}
    U^\upsilon = U^{\upsilon[\upphi]} 
    := \upsilon[\upphi]\- (U),
\end{align}
with $\upsilon\-$ the inverse map of $\upsilon$, s.t. $\upsilon \circ \upsilon\-=\id_M$. 
Crucially, these are $\Diff(M)$-invariant.
Indeed, defining the \emph{right action} of $\Diff(M)$ on subsets $U\subset M$ as $U \mapsto U^\psi:=\psi\- \circ U$ -- the action \eqref{Diff-trsf-fields} of $\Diff(M)$ on fields $\upphi$ being also a right action --  we find
\begin{align}
\label{inv-dressed-regions}
    (U^\upsilon)^\psi = (\upsilon[\upphi]^\psi)\- \circ (U^\psi) = \upsilon[\upphi]\- \circ \psi \circ \psi\- \circ (U) = U^\upsilon.
\end{align}
The reason for the definition \eqref{fielddepdrreg}, and the justification of the claim that dressed fields $\upphi^\upsilon$ live on such regions, comes from \emph{integration theory}. 
It tells us, e.g., that 
the action is invariant,
\begin{align}
    S:=\int_U L(\upphi) 
    = \int_{\,\psi\-(U)}\!\!\!\!\!\!\!\!\!\!\!\!\psi^*L(\upphi)
    = \int_{\,U^\psi} \!\!\!\! L(\upphi^\psi)
    =S^\psi,
\end{align}
where $L(\upphi)$ is the Lagrangian providing the dynamics of a general-relativistic theory, required to be $\Diff(M)$-covariant $L(\upphi^\psi)=\psi^*L(\upphi)$. 
By the DFM rule of thumb and \eqref{diffeodressedfields}, this gives
\begin{equation}
\label{int-dressed-fields}
S=\int_U L(\upphi) 
= \int_{\,\upsilon\-(U)}\!\!\!\!\!\!\!\!\!\!\!\! \upsilon^*L(\upphi)
=\int_{\,U^\upsilon}\!\!\!\! L(\upphi^\upsilon)
=:S^\upsilon.
\end{equation}  
One may observe that, upon applying the DFM, both terms in the integration pairing become separately invariant under diffeomorphisms.
The $\Diff(M)$-invariant regions $U^{\upsilon[\upphi]}$ thus represent \emph{physical} regions of spatiotemporal events,  
a physical spatiotemporal event being a field-dependent $\Diff(M)$-invariant point $x^{\upsilon[\upphi]}:= \upsilon[\upphi]\-(x) \in U^\upsilon$,  
realizing the basic intuition behind the point-coincidence argument.

The manifold $M$, like a scaffold for a building, is only there to
bootstrap our ability to build a description of the relativistic physics of fields' spatiotemporal d.o.f.  
And like
a scaffold, $M$ is removed from the physical picture by the  $\Diff(M)$-covariance of the general-relativistic field equations. 
On the other hand, we may now call $M^\upsilon := \left\{ U^\upsilon\, |\, \forall\ U\subset M\right\}$
the \emph{manifold of physical spatiotemporal events}.
Remark that it does not exist independently from the fields $\upphi$. 
The \emph{physical spacetime}, i.e., the physical Lorentzian manifold, is then  $(M^\upsilon,g^\upsilon)$. 
This technically implements the second of the dual insights stemming from  the point-coincidence argument, i.e., that physical spacetime 
is \emph{relationally defined}, in a $\Diff(M)$-invariant way, by its physical field content, and, in Einstein's own words, ``\emph{does not claim existence on its own,
but only as a structural quality of the field}''.

\subsection{Dressed dynamics of general-relativistic theories}\label{Dressed dynamics of general-relativistic theories}

The Lagrangian being $\Diff(M)$-covariant, so are the field equations $E(\upphi)=0$ derived from it. 
Given a dressing field $\upsilon$, the \emph{dressed Lagrangian}
\begin{align}
\label{dressed-Lagrangian-diffeo}
L(\upphi^\upsilon):=\upsilon^*L(\upphi)
\end{align}
is \emph{strictly} $\Diff(M)$-invariant by construction -- this is again a case of the DFM rule of thumb. 
The field equations for the dressed fields, $E(\upphi^\upsilon)=0$,
have the same \emph{functional} expression as the bare ones (for the bare theory) \cite{JTF-Ravera2024gRGFT}, but have a well-posed Cauchy problem, uniquely determining the evolution of the physical spatiotemporal d.o.f. represented by the dressed fields \eqref{diffeodressedfields}.
Again, the relational, dressed equations are the ones \emph{tacitly} used in practical experimental
situations. 
The fact that, as the result shows, they are ``equivalent'' to the bare field equations explains why GR/gRGFT was
successfully compared to experiments much before the problem
of identifying the fundamental physical d.o.f. and observables
was solved (cf. \cite{JTF-Ravera2024gRGFT} for details).

\subsection{There is no ``boundary problem'' in gRGFT}
\label{No boundary problem in gRGFT}

Gathering the above results and those of  
Section \ref{Dressing Field Method for gauge symmetries},
we obtain a $\big(\!\Diff(M)\ltimes \H)$-invariant and manifestly relational reformulation of gRGFT.
Defining a \emph{complete} field-dependent dressing field as the pair $(\upsilon, u)=(\upsilon[\upphi],  u[\upphi])$, we define, using \eqref{dressed-fields}-\eqref{diffeodressedfields}, the fully invariant dressed fields  
\begin{align}
    \upphi^{(\upsilon, u)} := \upsilon^*(\upphi^u),
\end{align}
representing the physical spatiotemporal and internal fields' d.o.f. and their
co-defining (coextensive) relations. 
The fields $\upphi^{(\upsilon, u)}$ live on/define the physical manifold of spatiotemporal events $M^{(\upsilon, u)}= M^\upsilon$, and regions $U^\upsilon$ thereof.\footnote{
Let us mention that the dressing field $(\upsilon,u)$ can be seen as the local version of a field-dependent dressing field $\bs u$ on a principal fiber bundle $P$ over $M$. The DFM procedure in this case allows to define the relational, physical enriched spacetime as a dressed bundle \cite{JTF-Ravera2024c,JTF-Ravera2024gRGFT,Berghofer:2025ius}.}
The dynamics is given by the fully invariant, dressed Lagrangian $L( \upphi^{(\upsilon, u)}):= \upsilon^*L(\upphi^u)$, 
from which we derive the field equations $E( \upphi^{(\upsilon, u)})=0$, with a well-posed Cauchy problem. 

An immediate consequence of all the above is that a physical, relationally-defined, boundary $\partial U^\upsilon$ is by definition $\big(\!\Diff(M)\ltimes \H\big)$-invariant.
This dissolves the so-called \emph{boundary problem} (see, e.g., \cite{DonnellyFreidel2016, Speranza2018,Geiller2018,Freidel-et-al2020-1,Kabel-Wieland2022,Ciambelli2023}), understood as the claim that
``$\Diff(M)$ and/or
$\H$ symmetries are broken at spacetime boundaries”.

Depending on the context and aims, various counter-measures are put forth in the literature to solve the boundary problem.
In the covariant phase space literature, this
involves, e.g., the \emph{ad hoc} introduction of so-called \emph{edge modes}, d.o.f. typically confined to $\partial U$, 
whose gauge transformations are tuned to cancel 
the terms from the transformation of the symplectic potential and 2-form, thereby  ``restoring their invariance'' \cite{DonnellyFreidel2016, Speranza2022,Geiller2017,Speranza2018, Geiller2018,Chandrasekaran_Speranza2021,Kabel-Wieland2022}.
Edge modes are sometimes interpreted as ``Goldstone modes'' resulting from the breaking of $\Diff(M)\ltimes \H$ at $\d U$.\footnote{Some have claimed that edge modes reveal essential new symmetries of gravity (so-called ``corner symmetries'') and may be key to a new paths to quantum gravity \cite{Freidel-et-al2020-1, Freidel-et-al2020-2, Freidel-et-al2020-3,Ciambelli2023}. Note also that edge modes can typically be reinterpreted as \emph{ad hoc} dressing fields, see, e.g., \cite{JTF-Ravera2024gRGFT}, imported from the outside into the theory to ``cure'' the boundary problem.} 
However, via the DFM, we can show that ``there is no boundary problem''.
In fact, if it is meant literally as above, it is based on a wrong statement. 
If it is meant as ``fields $\upphi$, or functional thereof, defined
at $\d U$ are not $\big(\!\Diff(M)\ltimes \H\big)$-invariant'' or as ``$\d U$ is not preserved by $\Diff(M)$'', it is technically true but trivial and physically inconsequential.

We observe that the boundary problem for GFT is easier to tackle/dissolve: it is sufficient to build $\H$-invariant field variables $\upphi^u$, since the manifold $M$ on which the fields $\upphi$ are defined does not transform under the action of $\H$.
The physical configuration of internal d.o.f. represented by $\upphi^u$ is $\H$-invariant across $M$, in particular at $\d U$ for $U\subset M$.

On the other hand,
the boundary problem in general-relativistic physics is more involved, since the manifold $M$ itself transforms under $\Diff(M)$. 
The dressing field $\upsilon[\upphi]$ is used to  dress both the bare fields $\upphi$ of the theory and  (regions $U$ of) $M$. 
The DFM for the case of diffeomorphisms encompasses diverse variants of  ``scalar coordinatization'' in GR. 
For instance, 
in the approach à la Kretschmann-Komar-Bergmann \cite{Komar:1958ymq, Bergmann-Komar1960, Bergmann1961,  Bergmann:1972ud} for vacuum GR, a dressing field is extracted from the bare metric, $\upsilon=\upsilon[g]$ (otherwise seen as a a ``$g$-dependent coordinate system''); 
the dressed metric $g^\upsilon:=\upsilon[g]^*g$ is ``self-dressed'', and it represents the invariant structure among the d.o.f. of the physical gravitational field. 
In approaches à la DeWitt \cite{DeWitt1960}, or Brown-Kucha\v{r} \cite{Brown:1994py, Rovelli:2001my},
if matter is described \emph{effectively} as a fluid (gas, particles, dust, etc.), it provides scalars $\upvarphi=\upvarphi^a$ 
from which one gets the dressing field $\upsilon[\upvarphi]$ (see, e.g., Section \ref{Scalar coordinatization of GR via the DFM});
the dressed metric $g^\upsilon:=\upsilon[\upvarphi]^*g$ represents the invariant relational structure between the d.o.f. of the metric and of the matter field --  otherwise seen as the metric ``written'' in the physical matter frame.

We refer the interested reader to \cite{JTF-RaveraNoBdyPb2025} for the simple construction for the case of GR, where, assuming that a dressing field $\upsilon$ for $\Diff(M)$ has been built from the field content $\upphi$ of the theory, it is proved explicitly the invariance of the dressed metric, i.e. of the physical gravitational field -- there, both the abstract and computationally concrete versions of the proof are provided (at the finite and infinitesimal/linearized levels).\footnote{We recall, for the sake of completeness, that diffeomorphisms $\psi \in \Diff(M)$ are smooth maps $\psi: M \to M$, $x' \mapsto \psi(x')$, with smooth inverse $x \mapsto \psi\-(x)$. 
Their linearization yields  vector fields $X := \tfrac{d}{d\tau}\psi_\tau\big|_{\tau=0}$, with $\psi_{\tau=0}=\id_M$,  which constitute the Lie algebra $\diff(M)\simeq \Gamma(TM)$.
As previously mentioned, the right action of $\Diff(M)$ on regions $U\subseteq M$ is defined as  is $U \mapsto \psi\-(U)$. 
The group $\Diff(M)$ acts on vector fields by \emph{pushforward}, $\psi_*: TM \to TM$, ${\mathfrak X}_{|x} \mapsto (\psi_*{\mathfrak X})_{|\psi(x)}$.
Its action on differential forms, and covariant tensors, defines the \emph{pullback}, $\psi^*: T^*M \to T^*M$, $\alpha_{|x}\mapsto (\psi^*\alpha)_{|\psi^{-1}(x)}$.
The latter is a also a right action. 
The linearisation of these actions defines the Lie derivative: 
on vector fields, $\L_X \mathfrak X := \tfrac{d}{d\tau} \psi_{\tau*} {\mathfrak X} \big|_{\tau=0} =[X, \mathfrak X]$;
on differential forms and covariant tensors, $\L_X \alpha := \tfrac{d}{d\tau} \psi^*_\tau \,\alpha\,\big|_{\tau=0}$.}

\subsection{Scalar coordinatization of GR via the DFM}\label{Scalar coordinatization of GR via the DFM}

We now illustrate the above formalism to the case of $D=4$ GR, with cosmological constant $\Lambda$, and matter described phenomenologically as a (perfect) fluid -- see \cite{JTF-RaveraNoBdyPb2025}.
The latter can be described by a set of scalar fields $\upvarphi=\upvarphi^a : U\subseteq M \rarrow N=\RR^4$, with $a=\{1,\ldots ,4\}$, characterizing the matter/fluid distribution and entering the expression of its covariantly conserved stress-energy tensor $\mathsf{T}=\mathsf{T}(g, \upvarphi)$,  $\nabla^g \mathsf T=0$.
This is itself derivable as the Hilbert stress-energy tensor of an effective Lagrangian $L_\text{{matter}}(g, \upvarphi)$ -- whose precise form is not needed here (cf. \cite{JTF-RaveraNoBdyPb2025} for details).

The fields $\upphi=\{g, \upvarphi\}$ $\Diff(M)$-transform as
\begin{align}
  \upphi^\psi
  =
  \{\psi^*g , \, \psi^*\upvarphi \}
  =
  \{ \psi^*g, \,\upvarphi \circ \psi \},
\end{align}
and the Lagrangian of the theory is 
\begin{equation}
\label{lagrangian-GR-scalar}
\begin{aligned}
L_\text{{GR}}(\upphi) &= L_\text{{GR}}(g, \upvarphi) \\
& = \tfrac{1}{2\kappa }\text{vol}_{g}  \big(\mathsf R(g) - 2 \Lambda \big) + L_\text{{matter}}(g, \upvarphi), \end{aligned}
\end{equation}
where $\kappa=\tfrac{8 \pi G}{c^4}$ is the gravitational coupling constant,  $\text{vol}_{g}$ is the volume 4-form induced by $g$, and $\mathsf R(g)$ is the  Ricci scalar.
The Lagrangian $L_\text{{GR}}(\upphi)$ is $\Diff(M)$-covariant, and so are the  Einstein equations derived from it: 
$E(\upphi)= \mathsf{G}(g) +\Lambda g  - \kappa \mathsf{T} (g, \upvarphi)=0$, 
where $\mathsf{G}(g)$ is the Einstein tensor. 
Let us remark that $\nabla^g\mathsf{T}=0$ implies that the fluid particles are in geodesic motion either if the pressure gradient vanishes, or if pressure itself does, in which case the fluid is a dust field (and we then make contact with \cite{Brown:1994py, Rovelli:2001my}).

We can identify a $\Diff(M)$-dressing field as 
\begin{align}
\upsilon=\upsilon[\upvarphi]:=\upvarphi^{-1}: \RR^4 \rarrow M,
\end{align}
as, indeed, $\upsilon^\psi=\upsilon[\upvarphi^\psi]=\psi\- \circ \upsilon[\upvarphi]$. 
This allows to define dressed regions
\begin{align}
    U^\upsilon := \upsilon\- (U), \quad \text{s.t.} \quad (U^\upsilon)^\psi = U^\upsilon.
\end{align}
In this way, we achieve a $\Diff(M)$-invariant relational definition of physical regions of events via (the scalar distribution of) matter 
as a \emph{physical reference  system}. 
The dressed fields are   
\begin{align}
     \upphi^\upsilon
     =
     \{g^\upsilon,\,  \upvarphi^\upsilon\} 
     =
     \{\upsilon^* g ,\,  \id_{U^\upsilon} \}.
\end{align}
Here, $\upvarphi^{\upsilon}:=$ {$\upsilon^* \upvarphi= \upvarphi \circ \upsilon = $} $\id_{U^\upsilon
}$ -- the matter distribution being ``self-dressed'' --
just expresses the fact that the (values of) the scalars now \emph{are} the coordinates, a.k.a. \emph{Lagrangian} or \emph{comoving} coordinates.
The $\Diff(M)$-invariant dressed metric $g^{\upsilon}$, encoding the geometric properties of $M^{\upsilon}$,   can then be understood as the physical gravitational field as measured in the coordinate system supplied by the matter distribution $\upvarphi$. 
In index notation, 
\begin{align}
\label{dressed-metric-index}
g^\upsilon_{ab} = {G(\upsilon)_a}^\mu \,  g_{\mu \nu} \, {G(\upsilon)^\nu}_b ,
\end{align}
with the Jacobian 
\begin{align}
G(\upsilon)=G(\upvarphi\-)=G(\upvarphi)\-\!=\left( \tfrac{\partial \upvarphi^a}{\partial x^{\,\mu}} \right).
\end{align}
The Lagrangian of the dressed, relational theory is
\begin{equation}
\label{Dressed-lagrangian-GR-scalar}
\begin{aligned}
L_\text{{GR}}(g^{\upsilon}, \upvarphi^{\upsilon}) 
:\!&=
{\upsilon}^*L_\text{{GR}}(g, \upvarphi)
 \\
& = \tfrac{1}{2\kappa }\text{vol}_{g^{\upsilon}}  \big(\mathsf R(g^{\upsilon}) - 2 \Lambda \big)  +   L_\text{{matter}}(g^{\upsilon}, \upvarphi^{\upsilon}). \notag
\end{aligned}
\end{equation}
From it, we derive the
\emph{relational Einstein equations},
\begin{align}
\label{dressed-Einstein-eq}
E(\upphi^\upsilon)=\mathsf{G}(g^{\upsilon}) +\Lambda g^{\upsilon} - \kappa \mathsf{T}(g^{\upsilon}, \upvarphi^{\upsilon})=0 ,
\end{align}
with $\mathsf{T} (g^\upsilon, \vphi^\upsilon)=:\mathsf{T}^\upsilon$ the conserved dressed stress-energy tensor, $\nabla^{g^\upsilon} T^\upsilon=0$ (controlling the dynamics of $\vphi^\upsilon$).
The relational Einstein equations\footnote{The dressed field equations have the same \emph{functional} expression as the bare ones. In components, they are related by
\begin{align*}
   \phantom{blabl}  \mathsf{G}^{\upsilon}_{ab} +\Lambda g^{\upsilon}_{ab} - \kappa \mathsf{T}^{\upsilon}_{ab} = {G(\upsilon)_a}^\mu \left( \mathsf{G}_{\mu \nu} + \Lambda  g_{\mu \nu} - \kappa  \mathsf{T}_{\mu \nu} \right) {G(\upsilon)^\nu}_b = 0.
\end{align*}
Note that this \emph{superficially} resembles the general-covariance of Einstein equations; yet, it is conceptually distinct.} 
are strictly $\Diff(M)$-invariant and have a well-posed Cauchy problem. 
Notice that, contrary to the bare equations, there is no more a ``pure metric side'' in the dressed field equations \eqref{dressed-Einstein-eq};  all the terms involve \emph{both} the physical metric and matter degrees of freedom.\footnote{To the attentive reader, this may naturally point toward a (\emph{relational}) quantization of gravity, insofar as the quantizable d.o.f. are now ``reshuffled'' with those of the metric -- cf. \cite{Francois:2025lqn}.}
These are the equations that are confronted to experimental tests \cite{JTF-Ravera2024gRGFT} (see, e.g., \cite{Francois:2025ptj} for an application to galaxy rotation curves analysis).

\section{Concluding remarks}\label{Concluding remarks}

These lecture notes have presented an introduction to symmetry reduction in gRGFT via DFM. 
The method provides a systematic and geometrically transparent procedure to construct gauge- and diffeomorphism-invariant variables -- so-called \emph{dressed fields} -- which encode the physical, \emph{relational} d.o.f. of the theory. 
Starting from the conceptual motivation rooted in the hole and point-coincidence arguments, the DFM was developed at both the kinematical and dynamical levels, highlighting its distinction from gauge fixing and its intrinsic relational interpretation.

A range of applications has been discussed across different physical settings. 
In gauge theories, the DFM allows one to construct invariant variables in models such as CS theory, Maxwell electromagnetism (including reinterpretations of gauge conditions as dressings), and the Abelian Higgs model without invoking SSB. 
Extensions to supersymmetric field theory show that fields commonly obtained via/studied with ``gauge fixing'' -- such as the RS field -- can instead be understood as self-dressed, invariant variables, with further developments leading to relational frameworks such as Matter–Interaction Supergeometric Framework.
Finally, in the general-relativistic context, the DFM provides a coherent treatment of diffeomorphism invariance, introducing dressed fields and dressed regions that implement a fully relational notion of spacetime, and resolve conceptual issues such as the hole argument and the boundary problem.

In its broader form, the DFM has been shown to apply to, e.g., gauge theoretic/Cartan geometric approaches to gravity \cite{Francois2014,AFL2016_I,Francois:2025jro}, conformal geometry and twistor theory \cite{Attard-Francois2016_I, Attard-Francois2016_II}, supersymmetric field theory \cite{JTF-Ravera2024-SUSY,JTF-Ravera2024ususyDFM,JTF-Ravera2025DFMSusyReview&Misu,JTF-Ravera2025offshellsusy}, general-relativistic physics \cite{Francois2023-a,JTF-RaveraNoBdyPb2025} and gRGFT \cite{JTF-Ravera2024gRGFT}, as well as quantum theory \cite{JTF-Ravera2024NRrelQM,Francois:2025lqn}.
The DFM has been shown to be the geometric underpinning of diverse efforts, spanning 15 years or so, to define invariant variables or build invariant constructions in various contexts.
For instance, since we have not discussed these in the present notes, let us recall the cases of the issue of the spin decomposition of nucleon \cite{FLM2015_I, Leader-Lorce}, ``edge modes'' à la Donnelly-Freidel dealing with apparent ``breaking'' of local symmetries at the boundaries (or corner) of compact regions \cite{DonnellyFreidel2016, Speranza2018, Teh-et-al2020, Teh-et-al2021} (on this claim, see \cite{JTF-RaveraNoBdyPb2025}), ``gravitational dressing'' à la Giddings-et-al \cite{Giddings-Donnelly2016-bis, Giddings-Kinsella2018, Giddings2019}, embedding maps/coordinates  \cite{Speranza2019} or dynamical reference frames \cite{Carrozza-Hoehn2021, Hoehn-et-al2022, Kabel-Wieland2022}, or yet, recently, ``intertwiner''/``dressing'' map \cite{Grassi:2024vkb}, 
as well as emerging applications beyond traditional field-theoretic settings in fundamental physics, such as symmetry-reduction-inspired approaches to relational structures in machine learning \cite{Francois:2026dvs}, etc.
All occurrences are unified,   technically and conceptually,  within the DFM framework.
No doubt many further examples can be found in the gRGFT literature.

Among the examples mentioned -- but not treated in detail -- in these lectures,
I would like to briefly point to further applications of the DFM to conformal gravity (cf. \cite{Attard-Francois2016_I} for details, and, e.g., the appendix of \cite{Andrianopoli:2026qsx} for a summary of the group-theoretic structure underlying conformal gravity) and to Metric-Affine Gravity (MAG).

Starting with conformal gravity, one can in fact achieve a complete reduction of the local symmetry group by constructing field variables that are manifestly invariant via the DFM. 
This is accomplished by first reducing the gauge symmetry associated with conformal boosts through the introduction of the dressing field 
\begin{align}
    q_a = {\sf{a}}_\mu \, {e^{\,\mu}}_a = {\sf{a}}_a,
\end{align}
which is a \emph{local} dressing involving the Weyl potential (with components ${\sf{a}}_\mu$). 
The resulting dressed theory still exhibits residual Lorentz and Weyl gauge symmetries - corresponding, in the DFM terminology, to residual transformations of the 1st kind. 
The Lorentz symmetry can then be reduced via
the tetrad as a Lorentz dressing field, as discussed in Section \ref{Lorentz dressing in GR}, 
while the Weyl symmetry -- being associated with \emph{group cocycles} (most naturally understood within the framework of cocyclic geometry) -- can be eliminated using tractor components as dressing fields (see \cite{Attard-Francois2016_I}). 
This completes the reduction of gauge symmetries in conformal gravity. 
One must then further reduce diffeomorphism invariance via the DFM, as shown in Section \ref{Dressing Field Method for general-relativistic theories}.

Finally, in the context of MAG, one is naturally led to confront the issue of so-called ``gauge translations'' $\T^n$. 
Let us recall that Cartan geometry provides the proper mathematical foundation for gauge theories of gravity, within which this issue does not arise, while still allowing for a clear geometric interpretation of translations. In particular, it has been shown in \cite{Francois:2025jro} that the kinematics of MAG can be systematically derived via the DFM in a technically streamlined way, thereby revealing that it effectively reduces to a Cartan-geometric kinematics.\footnote{The same reasoning can be applied to so-called ``Poincaré gauge gravity'', see \cite{Francois:2025jro}.}

More precisely, starting from a gauge theory based on the affine group 
$GL(n)\ltimes T^n$, with $n=\dim(M)$, the gauge potential naturally splits 
into a linear connection and additional fields associated with translations. 
Within the DFM framework, one can identify a $T^n$-valued dressing field $u$, 
which, in a compact matrix embedding, transforms as
\begin{align}
    u^{\mathbb{T}} = \mathbb{T}^{-1} u, \quad \mathbb{T}\in \T^n,
\end{align}
where $\T^n$ is the gauge group associated with $T^n$.
This dressing allows one to construct $\T^n$-invariant variables and thereby 
eliminate the translational gauge symmetry. 
In particular, the translational 
component of the gauge potential combines with the dressing field into a 
soldering form $e$, while the dressed connection reorganizes into the standard 
Cartan-geometric variables, namely a linear connection $A$ and a soldering 
form $e$. 
The corresponding dressed curvature similarly reproduces the 
Cartan-geometric field strengths. 
In this way, the MAG kinematics is seen 
to reduce to a Cartan-affine kinematics \cite{Francois:2025jro}.\footnote{In the DFM, the physical interpretation depends crucially on whether the 
dressing field is field-dependent or introduced independently. 
If it is 
added as an extra object to the bare kinematics, then it 
is an \emph{ad hoc} dressing field. 
In this form, in MAG, it reproduces the so-called 
``radius vector'' known in the MAG literature, which is introduced precisely 
to eliminate gauge translations $\T^n$. 
From the DFM viewpoint, this implies 
that in MAG (and similarly in Poincaré gauge gravity), the translational $\T^n$ sector is an \emph{artificial} (or ``fake'') gauge symmetry, with no independent physical signature, whereas only the residual $\mathcal{GL}(n)$ symmetry is substantive.

Alternatively, one may notice that the field content of MAG actually includes a $0$-form 
$\bar{X}$ (a local representative of a tensorial $0$-form on the 
$(GL(n)\ltimes T^n)$-bundle $Q\to M$), which transforms in the fundamental 
representation $R^n\simeq T^n$ of the affine group. 
In this case, a natural 
field-dependent $\T^n$-dressing field $u[\bar{X}]$ can be constructed. 
From the viewpoint of gauge theory, such fields are naturally interpreted 
as matter fields, or as ``Higgs'' fields when a potential term 
$V(\bar{X})$ is included in the Lagrangian. 
The resulting dressed variables 
can then be understood as relational, encoding the $\T^n$-invariant content 
of the theory, as discussed in \cite{Francois:2025jro}.}

Overall, both these lectures and the existing literature indicate that the DFM provides a coherent and unifying framework to systematically extract the relational, gauge- and diffeomorphism-invariant content of physical theories, thereby offering a clear pathway toward a deeper understanding of their fundamental d.o.f. in both classical and quantum contexts.

\section*{Acknowledgment}  %%%%%%%%%%%%%%%%%%%%%%%%%%%%%%%%%%%%%%%%%%

I take this opportunity to once again thank my colleagues V. Antonelli (DISMA, Politecnico di Torino, Italy) and J. François (University of Graz, Austria), co-organizers of the conference school \emph{Foundations of General-Relativistic Gauge Field Theory}, the latter also contributing as a lecturer in the module on bundle differential geometry in gRGFT, together with P. Berghofer (University of Graz, Austria) and M. Geiller (ENS Lyon, CNRS, France), for their insightful and engaging lectures, and all the participants who contributed to the success of the event through contributed talks and lively discussions.

L.R. is supported by the research grant PNRR Young Researchers, funded by MUR, MSCA Seal of Excellence (SoE), CUP E13C24003600006, ID SOE2024$\_$0000103, project GrIFOS.
This study was supported by the European Union - NextGenerationEU, under the Italian National Recovery and Resilience Plan (PNRR), Mission 4 Component 2 Investment 1.2, funding scheme ``Young Researchers'' (D.D. 201 del 3.7.2024). 
This manuscript reflects only the author's views and opinions, and the Ministry cannot be considered responsible for them.

\appendix

\section{``Stueckelberg trick'' \emph{vs} the DFM}\label{Stueckelberg trick vs the DFM}

In this appendix, we review the technical and conceptual differences between the Stueckelberg trick and the DFM. 
To this end, we consider the simple case of a GFT with field content $\Phi = \{A\}$, where $A = A_\mu \, dx^\mu$ is a vector field (1-form potential), and take the gauge group to be $\mathcal{H} = \U(1)$.
Let us first consider what happens with the Stueckelberg trick, starting from a Proca Lagrangian (for simplicity, the Abelian model) for $A$, through which the $\U(1)$ symmetry is \emph{implemented} (i.e., \emph{introduced} in the theory), and then turn to the construction within the DFM, in which the symmetry is instead \emph{reduced}.

\subsection{Stueckelberg trick, to \emph{implement} a symmetry}

We consider the Proca theory given by the Lagrangian
\begin{align}
    L_{\text{Proca}}(A) = \tfrac{1}{2} F\star F + m^2 A \star A ,
\end{align}
which has a massive $A$, and no $\U(1)$ symmetry.
We then introduce, by hand, a Stueckelberg field $B$ s.t.
\begin{align}
\label{Btr}
    B^\gamma = B-\lambda , \quad \gamma= e^\lambda \in \U(1) ,
\end{align}
and we require that $A$ transforms as
\begin{align}
    A^\gamma = A + d \lambda .
\end{align}
We add $B$-terms in the Lagrangian, s.t.
\begin{align}
\label{LStueckelberg}
    L_{\text{Stueckelberg}} (A,B) = \tfrac{1}{2} F\star F + m^2 (A+dB) \star (A +dB).
\end{align}
The Lagrangian \eqref{LStueckelberg} now enjoys the $\U(1)$ symmetry.
We have therefore implemented the $\U(1)$ gauge symmetry by introducing into the theory the (extra) Stueckelberg field $B$.
From \eqref{Btr}, we see that the Stueckelberg field formally transforms as a $\U(1)$ dressing field, and we may therefore call it an \emph{ad hoc} dressing field.
However, as we shall see below, the Stueckelberg trick and the DFM are conceptually different, as the former is used to implement the gauge symmetry, while the latter is used to reduce it, building gauge-invariant objects.

\subsection{DFM, to \emph{reduce} a symmetry}

With the DFM, we start from the theory $L_{\text{Stueckelberg}} (A,B)$, with $\upphi= \lbrace{ A,B\rbrace}$, enjoying the $\U(1)$ symmetry, and reduce the latter.
The dressing field is
\begin{align}
    u(\upphi) = u (B)=e^B ,
\end{align}
s.t.
\begin{align}
    u(\upphi)^\gamma = u (\upphi^\gamma) = e^{B^\gamma} = e^{B-\lambda} = \gamma^{-1} u (\upphi).
\end{align}
The dressed fields are then (in this simple Abelian case)
\begin{align}
    A^u = A + u^{-1} du = A+dB , \quad F^u=F ,
\end{align}
and
\begin{align}
\label{invL}
    L_{\text{Stueckelberg}} = L_{\text{Proca}} (A^u) = \tfrac{1}{2} F^u \star F^u + m^2 A^u \star A^u = \tfrac{1}{2} F \star F + m^2 A^u \star A^u .
\end{align}
Remark that there is no $\U(1)$ symmetry in the dressed theory given by the dressed Lagrangian \eqref{invL}.

\section{Dirac dressing as an instance of the DFM}\label{Dirac dressing}

As a further example of the application of the DFM in GFT, we may consider the ``Dirac dressing'' (1958) in EM (setting all fundamental constants to $1$), with Abelian gauge symmetry $\U(1)$. 
The field content of the theory is given by a spinor $\psi$ (representing the electron) and the EM gauge potential $A = (A_0, A_r)$.

Dirac observed that gauge variables are subject to a non-deterministic evolution, which raises problems in quantization. 
He therefore proposed to consider new variables
\begin{align}
    \psi^* = e^{iC}\psi , \quad A^* = A + dC,
\end{align}
and thus the covariant derivative
\begin{align}
d\psi^* - i A^* \psi^* = e^{iC}(d\psi - i A \psi),
\end{align}
where the phase factor $C=C(x)$ is 
\begin{align}
    C(x) = \int c_r (x,x') A^r (x') d^3x' .
\end{align}
For $\psi^*$ and $A^*$ gauge-invariant, 
\begin{align}
    \tfrac{\partial}{\partial x^i_r} c_r(x,x')=\delta (x-x') ,
\end{align}
which admits the Coulomb potential as a solution.
When quantizing, $\psi^*=\psi^*(x)$ was interpreted by Dirac as an ``operator of creation of an electron together with its Coulomb field'', which is a physical variable.

This is an instance of the DFM.
Indeed, under gauge transformations of the gauge potential, the phase factor transforms as
\begin{equation}
\begin{aligned}
    C'(x) & = \int c_r (x,x') {A'}^r (x,x') d^3 x' = C(x) + \int c_r (x,x') \tfrac{\partial S}{\partial x'_r} (x') d^3x' = C(x) \\
    & \phantom{=} - \int \tfrac{\partial}{\partial x'_r} c_r (x,x') S(x') d^3 x'\\
    & = C(x)- S(x).
\end{aligned}
\end{equation}
So,
\begin{align}
\label{nonlocdiracdr}
    u=e^{iC}
\end{align}
transforms under $\gamma=e^{iS} \in \U(1)$ as
\begin{align}
    u' = \gamma^{-1} u ,
\end{align}
that is, it is an Abelian dressing field.
Hence, $\psi^*$ and $A^*$ in Dirac's equations are Abelian instances of dressed fields.
Note that $u$ is \eqref{nonlocdiracdr} is a non-local dressing field, signaling that the reduced $\U(1)$ symmetry is substantive \cite{Francois2018} in this model.

\bibliographystyle{JHEP}
\bibliography{bibliolect}

\end{document}